\documentclass[a4paper,fleqn,usenatbib]{mnras}

\usepackage{newtxtext,newtxmath}

\usepackage[T1]{fontenc}
\usepackage{ae,aecompl}
\usepackage[utf8]{inputenc}

\usepackage{graphicx}	
\usepackage{amsmath}	
\usepackage{lscape}
\usepackage{longtable}

\usepackage[english]{babel}
\newcommand{\myr}{$\rm mas\,yr^{-1}$}
\newcommand{\kms}{$\rm km\,s^{-1}$}
\newcommand{\kmskpc}{$\rm km\,s^{-1}\,kpc^{-1}$}

\title[Kinematics of the Milky Way from the Gaia EDR3 data] {Kinematics of the Milky Way from the Gaia EDR3 red giants and sub-giants.}

\author[Fedorov et al.]{
P. N. Fedorov,$^{1}$\thanks{E-mail: pnfedorov@gmail.com (PNF)}
V. S. Akhmetov,$^{1}$
A. B. Velichko,$^{1}$
A. M. Dmytrenko$^{1}$  \newauthor 
and S. I. Denischenko$^{1}$
\\
$^{1}$Institute of astronomy of V. N. Karazin Kharkiv national university, Svobody sq. 4, 61022, Kharkiv, Ukraine\\
}

\date{Accepted XXX. Received YYY; in original form ZZZ}

\pubyear{2021}

\begin{document}
\label{firstpage}
\pagerange{\pageref{firstpage}--\pageref{lastpage}}
\maketitle

\begin{abstract}
We present the results of the kinematic investigations carried out with the use of spatial velocities of red giants and sub-giants containing in the $Gaia$~EDR3 catalogue. The twelve kinematic parameters of the Ogorodnikov--Milne model have been derived for stellar systems with radii 0.5 and 1.0 kpc, located along the direction the Galactic center -- the Sun -- the Galactic anticenter within the range of Galactocentric distances $R$ 0--8--16~kpc. By combining some of the local parameters the information related to the Galaxy as a whole has been received in the distance range 4--12~kpc, in particular the Galactic rotational curve, its slope, velocity gradients. We show that when using this approach, there is an alternative possibility to infer the behaviour of the Galactic rotational curve and its slope without using the Galactocentric distance $R_\odot$. The kinematic parameters derived within the Solar vicinity of 1~kpc radius are in good agreement with those given in literature.
\end{abstract}


\section{Introduction}
\label{sub:intro}

Investigation of stellar kinematics is an important instrument for the Galactic study. It is based on the analysis of stellar proper motions and radial velocities within various physical and mathematical models. The Ogorodnikov--Milne (O--M, \cite{Ogorodnikov1932, Milne1935, Ogorodnikov1965}) model is most commonly used for the stellar kinematics analysis. The model provides the ability to investigate the velocity field in a deformable stellar system. The analysis of the stellar velocity field is usually confined to solving the basic kinematic equations to compute the O--M model parameters in the local coordinate system, which moves together with the Sun about the Galactic centre.

In the last time, some new models are often used to analyse the stellar kinematics. Since stellar proper motions and radial velocities are the components of the velocity vector field, it would be reasonable to use the methods of decomposition of the corresponding stellar velocity field on a set of the vector spherical harmonics (VSH). The approach allows to detect all the systematic constituents present in the stellar velocity field under study, as well as make the mathematically complete kinematic model. Comparison of the decomposition coefficients with model parameters shows whether the models are complete as well as allows to reveal all significant systematics which is not included into the models. This approach is effective, as demonstrated in a number of papers, for instance \cite{Vityazev2004, Vityazev2005, Makarov2007,  Mignard2012, Velichko2020}.

\section{The O--M model equations and forming the local coordinate systems}
\label{sec:scheme}
 
\begin{figure}
   \centering
\resizebox{\hsize}{!}
   {\includegraphics{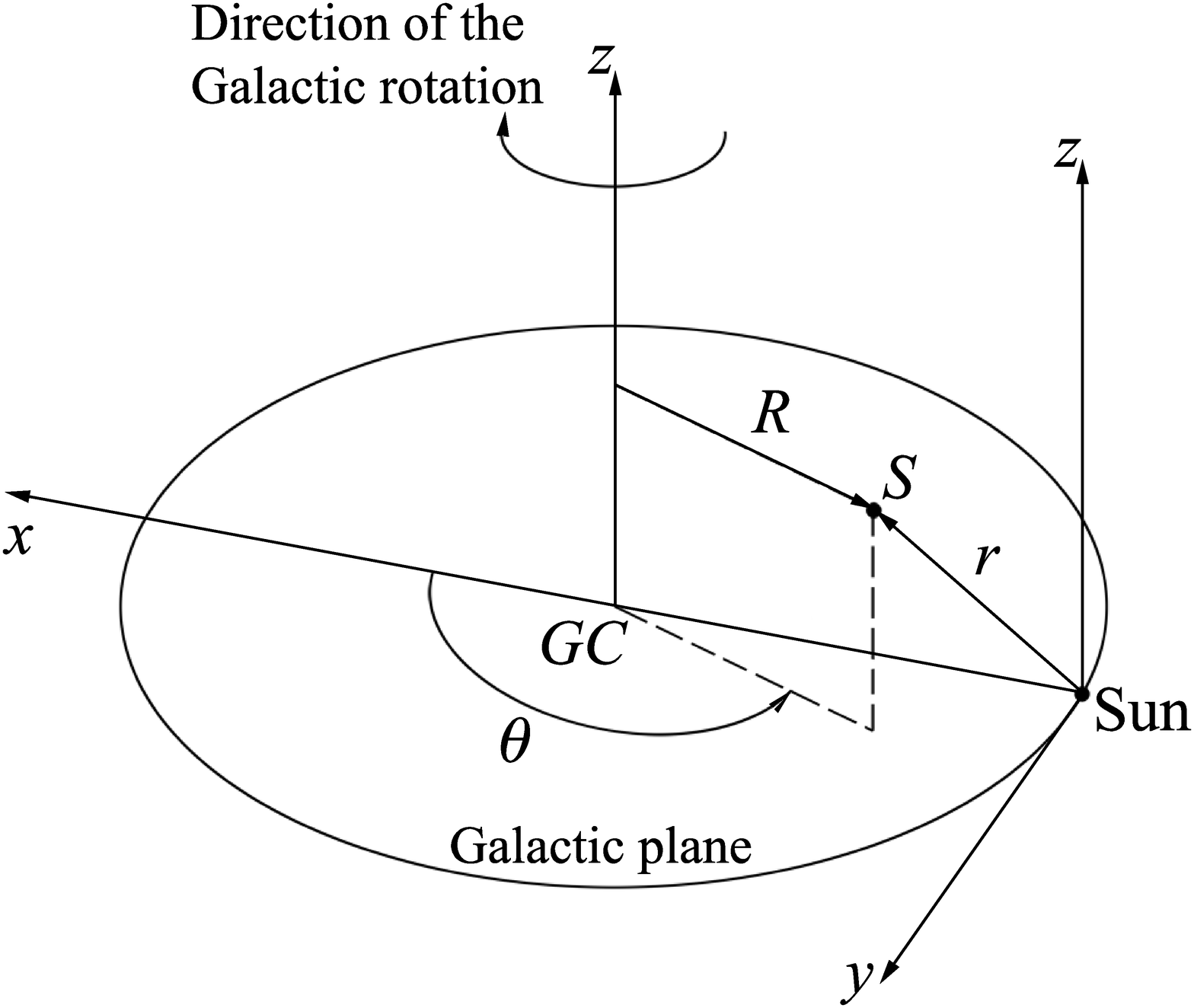}}
   \caption{The local rectangular Galactic $x, y, z$ and cylindrical Galactocentric $R, \theta, z$ coordinate systems. $S$ is an arbitrary star with the heliocentric distance $r$.}
\label{fig:GCS}%
\end{figure}

To write mathematically the basic kinematic equations of various models, the Cartesian coordinate system $x, y, z$ is usually applied. The origin of the system coincides with the Solar system barycenter. The $x$ axis points to the Galactic center, $y$ axis coincides with the direction of the Galactic rotation, while $z$ axis is perpendicular to the Galactic plane and complements the right-handed Cartesian coordinate system (see Fig. \ref{fig:GCS}). The given system is called the local rectangular Galactic coordinate system.

In this paper, we use the O--M model which describes the systematic differential stellar velocity field within the vicinity of a chosen point. According to the Helmholtz theorem, the stellar motion may be represented by the sum of velocities, namely the translational motion of the centroid of the corresponding stellar system, as well as its rigid-body rotation and deformation velocities. By the term "centroid" we mean a point that coincides with that moving at a velocity equal to the mean velocity of stars belonging to the given stellar system. This approach has been used, for instance, in the papers by \cite{Clube1972, duMont1977, Miyamoto1993, Miyamoto1998, Bobylev2011}.

In general, representation of the velocity field of a deformable stellar system is based on the general form of expanding the continuous vector function $\bf V(r)$ which is defined in the vicinity of any point $\bf r_0$ ($x_0, y_0, z_0$). In the Cartesian Galactic coordinate system $x, y, z$ the general form of the stellar velocity field expansion $\bf V(\bf r)$ is as follows:

\begin{equation}
\label{eq:stel_f}
{\bf V}({\bf r}) = {\bf V}({\bf r}_0 +  d{\bf r}) = {\bf V}({\bf r}_0) + \left( \frac{\partial V}{\partial x_k} \right)_0 dx_k + \frac{1}{2!}\left(\frac{\partial^2V}{\partial x_k \partial x_m}\right)_0dx_kdx_m + ...
\end{equation}
where the indexes $k, m = $  1,2,3, and the partial derivatives are evaluated at the point  ${\bf r}_0(x_0, y_0, z_0)$.

When analysing the solar vicinity, we restrict the expansion \ref{eq:stel_f} to the first-order terms and derive the components $v_i$ of the linear velocity field $\bf V$ at the point located at the heliocentric distance from the Sun
${\bf r} = {\bf r}_0 + d{\bf r}$ as follows:

\begin{eqnarray}
\label{eq:sol_m}
&& v_i({\bf r}_0 + d{\bf r}) =  v_i({\bf r}_0) + \frac{1}{2}\left( \frac{\partial v_i}{\partial x_k} - \frac{\partial v_k}{\partial x_i}\right)_0dx_k + \\ \nonumber
&& + \frac{1}{2}\left( \frac{\partial v_i}{\partial x_k} + \frac{\partial v_k}{\partial x_i}\right)_0dx_k = v_i({\bf r}_0) + M^-dx_k + M^+ dx_k
\end{eqnarray}
where $i, k$ = 1, 2, 3.

In this equation, $v_i({\bf r}_0)$ is usually interpreted as the mean velocity of the stellar system relative to the Sun, or on the contrary, as the Solar velocity relative to the chosen centroid with the opposite signs, where the components are $v_i({\bf r}_0) = -X_\odot {\bf e}_x -Y_\odot {\bf e}_y -Z_\odot {\bf e}_z$. Thus, finally one can write:

\begin{eqnarray}
\label{eq:sol_m}
v_i(r_0 + dr) = - X_\odot {\bf e}_x - Y_\odot {\bf e}_y - Z_\odot {\bf e}_z +  M^-dx_k + M^+ dx_k     
\end{eqnarray}

The matrix elements $M^-$ and $M^+$ are the partial derivatives of the projections $(v_1, v_2, v_3)$ of the velocity
vector on the axes of the rectangular Galactic coordinate system, and they are usually called kinematic parameters.

Components of the antisymmetric matrix
\begin{equation}
M^- = \frac{1}{2}\left( \frac{\partial v_i}{\partial x_k} - \frac{\partial v_k}{\partial x_i}\right)_0 = \omega_{ik} = -\omega_{ki}
\end{equation}
constitute antisymmetric tensor which is referred to as the tensor of local rotation velocities. It describes the
rigid-body rotation of the stellar system under examination about the axis which passes through the stellar centroid
with the instant angular velocity ${\bf \omega}$, and $\omega_1, \omega_2, \omega_3$ are projections of ${\bf \omega}$
on the Galactic axes $x, y, z$ respectively.

The matrix
\begin{equation}
M^+ = \frac{1}{2}\left( \frac{\partial v_i}{\partial x_k} + \frac{\partial v_k}{\partial x_i}\right)_0 = u_{ik}
\end{equation}
is symmetric, and its components $v_{ik} = v_{ki}$ form the symmetric second-rank tensor which is referred to 
as the tensor of local deformation velocities. It defines the velocity of deformation motion in the stellar system under study. Usually, it is these components of the tensor that correspond to the forces inducing deformations in the dynamical investigations (\cite{Tarapov2002, Sedov1970}).

The diagonal components of the symmetric matrix $M^+_{11}$, $M^+_{22}$ and $M^+_{33}$ characterize velocities of relative contractions/expansions of the stellar system along $x, y, z$ axes, while the components $M^+_{12} = M^+_{21}$, $M^+_{23} = M^+_{32}$, $M^+_{13} = M^+_{31}$ do velocities of the angular deformations in the $(x, y)$, $(y, z)$ and $(x, z)$ planes, respectively. Here the velocity of angular deformation means the velocity of changing angles between segments in the planes.

To set a relation between the velocity components ${\bf V}({\bf r})$ and the observational data, namely proper motions $\mu_l$ and $\mu_b$ on the Galactic longitude and latitude, respectively, as well as radial velocity $V_r$ of a star, we project the vector ${\bf V(r)}$ on the unit vectors of the spherical Galactic coordinate system ${\bf e}_l$, ${\bf e}_b$, ${\bf e}_r$ and introduce the factor $k = 4.74057$ to transform the dimension of stellar proper motions from \myr\, to \kmskpc. Then, the conditional equations for proper motions and radial velocities in the local Galactic coordinate system have the following form:

\begin{align}
\ k\,\mu_l\,{\rm cos}\,b &= X_\odot/r\,{\rm sin}\, l - Y_\odot/r\, {\rm cos}\,l - \omega_1\,{\rm sin}\,b\,{\rm cos}\, l  - \omega_2\,{\rm sin}\,b\,{\rm sin}\,l + \nonumber  \\
& + \omega_3\,{\rm cos}\,b\, + M^+_{12}\,{\rm cos}\,b\,{\rm cos}\,2l - M^+_{13}\,{\rm sin}\,b\,{\rm sin}\, l  + \nonumber \\
&+ M^+_{23}\,{\rm sin}\,b\,{\rm cos}\,l
 - 0.5\,M^+_{11}{\rm cos}\,b\,{\rm sin}\,2l + \nonumber \\
 & + 0.5\,M^+_{22}{\rm cos}\,b\,{\rm sin}\,2l  \label{eq:OMM_mul}
 \end{align}
 \begin{align}
  \ k\,\mu_b & = X_\odot/r\,{\rm cos}\, l\,{\rm sin}\,b + Y_\odot/r\,{\rm sin}\, l\,{\rm sin}\, b\, - Z_\odot/r\,{\rm cos}\, b  + \omega_1\,{\rm sin}\, l - \nonumber \\
 & - \omega_2\,{\rm cos}\, l -  0.5\, M^+_{12}\,{\rm sin}\, 2b\,{\rm sin}\, 2l + M^+_{13}\,{\rm cos}\, 2b\,{\rm cos}\,l + \nonumber \\
 & + M^+_{23}\,{\rm cos}\, 2b \,{\rm sin}\, l\, - 0.5\,M^+_{11}\,{\rm sin}\,2b\,{\rm cos}^2\, l - \nonumber \\ 
 & - 0.5\, M^+_{22}\,{\rm sin}\,2b\,{\rm sin}^2\,l + 0.5\,M^+_{33}\,{\rm sin}\,2b  \label{eq:OMM_mub}
 \end{align}
 \begin{align}
 V_r/r & = -X_\odot/r\,{\rm cos}\,l{\rm cos}\,b - Y_\odot/r\,{\rm sin}\,l\,{\rm cos}\,b - Z_\odot/r\,{\rm sin}\,b + \nonumber \\
 & + M^+_{13}\,{\rm sin}\,2b\,{\rm cos}\,l + M^+_{23}\,{\rm sin}\,2b\,{\rm sin}\,l + M^+_{12}\,{\rm cos}^2\,b\,{\rm sin}\,2l + \nonumber \\
 & + M^+_{11}\,{\rm cos}^2\,b\,{\rm cos}^2\,l  + M^+_{22}\,{\rm cos}^2\,b\,{\rm sin}^2\,l + M^+_{33}\,{\rm sin}^2\,b  \label{eq:OMM_vr}   
\end{align}

The equations given above define the differential stellar velocity field and contain 12 unknown parameters.

The local Galactic coordinate system can be introduced at any arbitrary point of the Galactic plane provided that one know coordinates $x, y, z$,  spacial velocities  $V_x, V_y, V_z$ of the point (for instance, a star) and for all stars located within the corresponding vicinity. Taking the Galactocentric distance of the Sun to be equal to $R_\odot = 8.0$ kpc (\cite{Vallee2017}), one can make the transition from the local Galactic coordinate system with the origin at the barycenter of the Solar system to one with the origin at the chosen point. The corresponding transformation takes into account, in general, transposing the origin of the Galactic coordinate system and turning the coordinate axes. The $x'$--axis of the new coordinate system always points to the Galactic center, the $y'$--axis coincides with the
direction of the Galactic rotation, and the $z'$ one is perpendicular to the Galactic plane. The procedure would translocate the fictitious observer from the barycenter of the Solar system to the point given by the origin of the new coordinate system.

Thus, one can write the conditional equations of the O--M model in the chosen coordinate system and estimate the kinematic parameters in the vicinity of each chosen point. It is evident that these kinematic parameters are local, since they characterize the stellar kinematics within a small spacial volume, for instance within a spherical area with a given radius $r'$. Nevertheless, such an analysis of stellar kinematics in various Galactic parts provides insight into behaviour of some global kinematic parameters, i.e. those related to the entire Galaxy.

In this paper, we present estimations within the O--M model of the kinematic parameters for stellar systems located along the direction the Galactic center -- the Sun -- the Galactic anticenter. In this case, the origin of the coordinate system is shifting along the $X$-axis relative to the Sun with the step of 250~pc, without any rotations of the coordinate system.

\section{Solving the problem}

\begin{figure}
   \centering
\resizebox{\hsize}{!}
   {\includegraphics{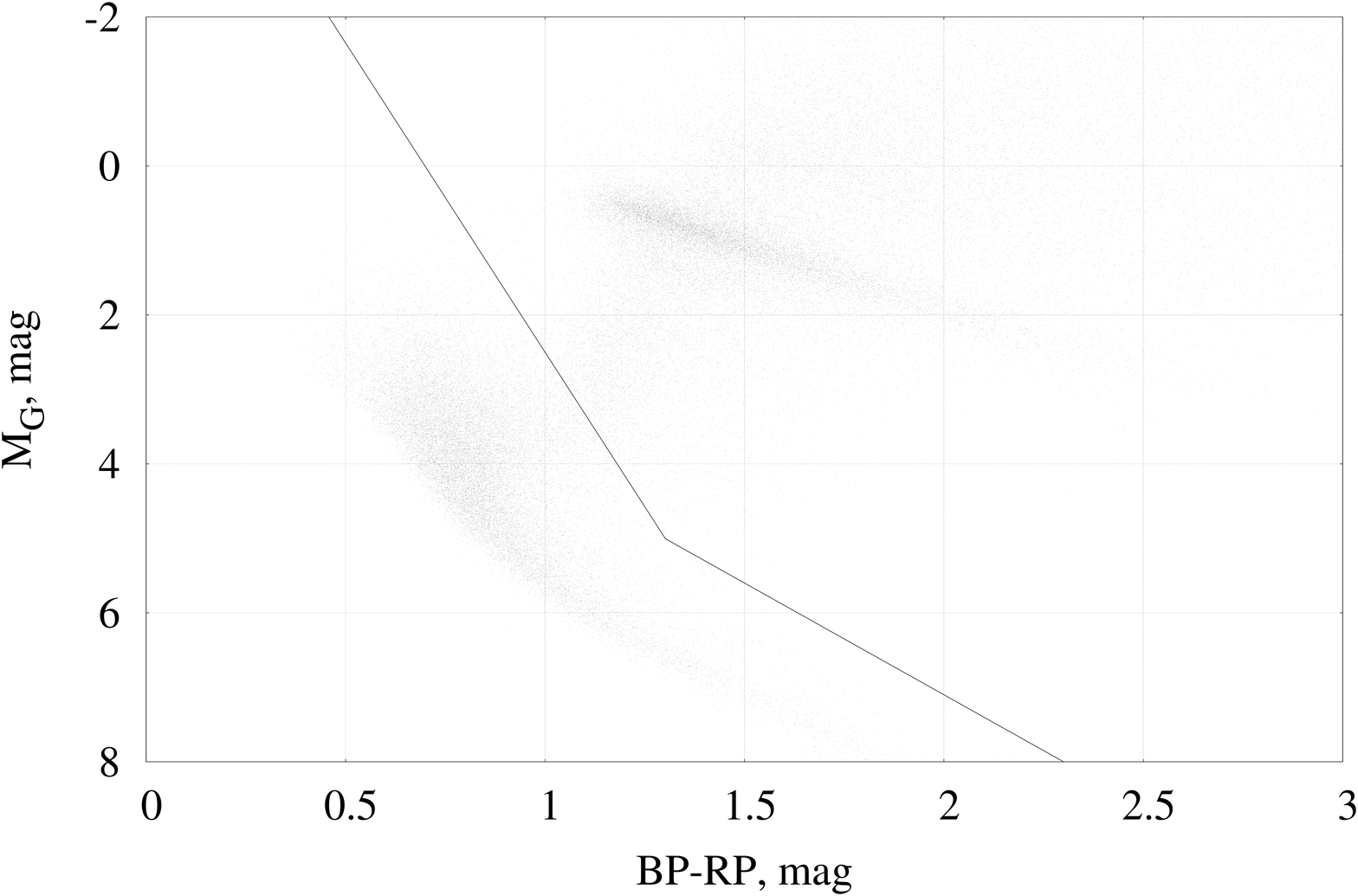}}
  \caption{Selection of red giants and sub-giants.}
\label{fig:MS_RG}%
\end{figure}

For the analysis, we use the $Gaia$~EDR3 data (\cite{Brown2016, Brown2021}) of $\sim$7,21 million stars, for which, in addition to proper motions, radial velocities were also derived within the second release of the $Gaia$ mission (\cite{Katz2019}. The diagram color $BP-RP$ -- absolute magnitude $M_G$ for 10\% of these stars is shown in Fig.~\ref{fig:MS_RG}. The diagram is rough because extinction was not be taken into account. Using the straight lines in Fig.~\ref{fig:MS_RG}, the stars were split into the main sequence stars (located below the lines), and higher luminosity stars, such as sub-giants and giants (located above the lines)

\begin{figure}
   \centering
\resizebox{\hsize}{!}
   {\includegraphics{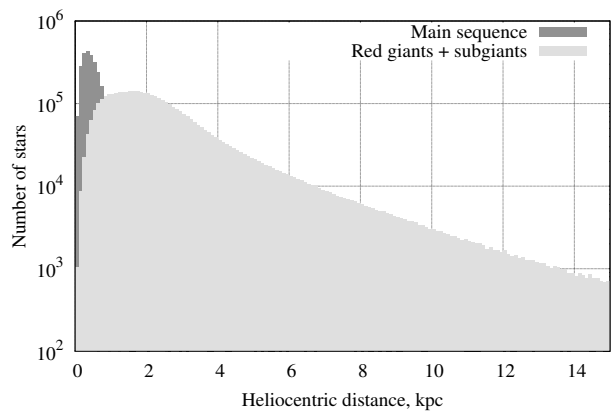}}
  \caption{The number of main sequence stars and red giants plus sub-giants depending on the heliocentric distance.}
\label{fig:Star_numbers}%
\end{figure}

The Fig.~\ref{fig:Star_numbers} shows the distribution of number of stars with the heliocentric distance $r$. $r$ were computed from parallaxes as $1/\pi$. As one can see from the Fig.~\ref{fig:Star_numbers}, the distance range corresponding to the main sequence stars occupies the relatively narrow range of the Galactocentric distances 7 kpc $< R < $9 kpc. On the contrary, sub-giants and red giants cover a much wider Galactocentric distance range, 0 kpc~$< R <$~16 kpc. Therefore, the latter stellar sample is better to trace the Galactic kinematics. The total amount of sub-giants and red giants is 4.5~million.

Each point, from which the observations would be made by a fictitious observer, is placed on the $x$ axis. The centers of the spheres, restricting spacial volumes that contain stars around each point, are located every 250~pc from $R =$0~kpc to 16~kpc . The radii $r'$ of the spheres have been set equal to 500~pc or 1000~pc.

To improve the accuracy of the derived results, in the specific coordinate system the special procedure has been carried out consisting in exclusion stars whose velocities deviate from the average value by more than 3$\sigma$. The rejection procedure has been carried out in each pixel on the celestial sphere that would be visible from the specific point. The pixelization has been made according to the HEALPix scheme (The Hierarchical Equal Area iso-Latitude Pixalization, \cite{Gorsky2005}) with N$_{\rm side}=$~10 that allows to divide the sphere onto 1200 equal areas (pixels). The pixelization has been applied only for excluding stars.

Once the rejection procedure has been applied, the individual values of stellar positions, parallaxes, proper motions,
and radial velocities has been used to solve by the linear least square method (LSM) the O--M model equations written for each local Galactic coordinate system. As a result, the derived kinematic parameters characterize the stellar velocity field within vicinity of each given point.

The O--M model parameters allow to estimate a number of kinematic and physical parameters related to the Galaxy as a whole. Within the simplified form of the O--M model is usually called as the Oort-Lindblad one \cite{Ogorodnikov1965}, the stellar velocity field is assumed to be axisymmetric. Herewith, the system rotates in the Galactic plane only, i.e. $\omega_3$ = $|\bf \omega|$ while $\omega_1$ and $\omega_2$ are equal to zero. In this case the value of $M^-_{21}$ = $\omega_3$ corresponds to the Oort constant $ B $. The deformation also exists in the Galactic plane only, i.e. it is characterized by the Oort constant $A = M^+_{12}$, while  $M^+_{13}$ and $M^+_{23}$ are equal to zero. Contractions/expansions  $M^+_{11}, M^+_{22}, M^+_{33}$ along the corresponding axes are not taken into account. The Oort constants are usually used in papers on kinematics to characterize the Galactic rotation (for instance, \cite{Bovy2017, Vityazev2018, Chengdong2019, Tsvetkov2019}). Thus, the angular rotation velocity of the Galaxy in the Solar vicinity is equal to

\begin{equation}
\Omega = A - B = M^+_{12} - \omega_3
\end{equation}

In general, if not to restrict the stellar velocity field, taking into account the fact that the pairs ($\omega_1$, $M^+_{23}$) and ($\omega_2$, $M^+_{13}$) in the corresponding planes are analogous to the Oort constants $B$ and $A$ respectively, for an arbitrary centroid located at the Galactocentric distance $R$ one can compute the angular rotation of the Galaxy velocity $\Omega_R$:

\begin{equation}
\Omega_R = \sqrt{(M^+_{12} - \omega_3)^2+(M^+_{13} - \omega_2)^2+(M^+_{23} - \omega_1)^2}
\label{eq:Omega_R}
\end{equation}

It is obvious that, knowing values of the Solar velocity components $X_\odot, Y_\odot$ and $Z_\odot$, or velocity of an arbitrary point where the origin of the coordinate system is placed, relative to the centroids, one can compute the velocity module $V_\odot$ and the apex coordinates $L_\odot$ and $B_\odot$:

\begin{equation}
V_\odot = \sqrt{X^2_\odot+Y^2_\odot+Z^2_\odot}
\end{equation}

\begin{equation}
{\rm tg}\,L_\odot = \frac{Y_\odot}{X_\odot},~~~
{\rm tg}\,B_\odot = \frac{Z_\odot}{\sqrt{X^2_\odot + Y^2_\odot}} 
\end{equation} 

\section{Kinematic parameters of the O--M model}

It is worth noting that, at this stage of our kinematic analysis the O--M model parameters derived out of the range of Galactocentric distances 4--12 kpc, are not assumed to be reliable due to a variety of reasons. This is, first of all, significant decreasing the amount of stars (the number of stars containing in each stellar system is given in tables \ref{tab:model_part1},\ref{tab:model_part2}). As a result, the uncertainties of the derived parameters increase. Moreover, the error bars grow since the astrometric parameters become less accurate with the heliocentric distance to the sources. In third place, since the O--M model is linear, it can be confidently applied if the radius of the stellar system under examination $r'$ is much less than the Galactocentric distance to the stellar system $R$, i.e. $r'<<R$. Unfortunately, this condition is not fulfilled when analysing the stellar systems which are close to the Galactic center ($R<$~4~kpc), although toward the Galactic anticenter the condition if well fulfilled. 

However, the main reason of unreliability is due to our estimations of the parameters of rotation $\omega_1, \omega_2$ and deformation in $yz$ and $xz$ planes are actually the values averaged over the northern and southern Galactic hemispheres. In the papers by \cite{Vityazev2014, Velichko2020} the stellar kinematics was analysed using the vector spherical harmonics. The authors have shown that the values of the deformation velocity component $ M^+_{12}$ in the $xy$ plane and rotation velocity $\omega_3$ are equal in the northern and southern Galactic hemispheres. The values of the rotation velocity components $\omega_1$, $\omega_2$ as well as deformation velocity components $ M^+_{23}$ and $ M^+_{13}$ have different signs in the northern and southern hemispheres while their modules are virtually equal. Therefore, it is not surprising that we see in Fig.~\ref{fig:OMM_par} the near zero averaged values of these parameters within the range 4~$< R <$~12~kpc, since the centers of the spheres with radii $r'$ 500~pc and 1000~pc are located in the Galactic plane and encompass the stars of both the northern and southern Galactic hemispheres. However, it would be wrongly to consider as actual any deviations of the values from zero out of the Galactocentric range 4--12~kpc, since the deviations can be caused, for instance, by shifting the LSM solutions due to the celestial sphere is not covered uniformly (distribution of objects through the whole celestial sphere is not uniform). Nevertheless, we show in figures the Galactocentric distance range from 0~kpc to 16~kpc to have at least some understanding of behaviour of the kinematic parameters near the Galactic center as well as in its outer part.

\subsection{Components of the rotational tensor}
\label{sec:omega}

\begin{figure*}
   \centering
\resizebox{\hsize}{!}
   {\includegraphics{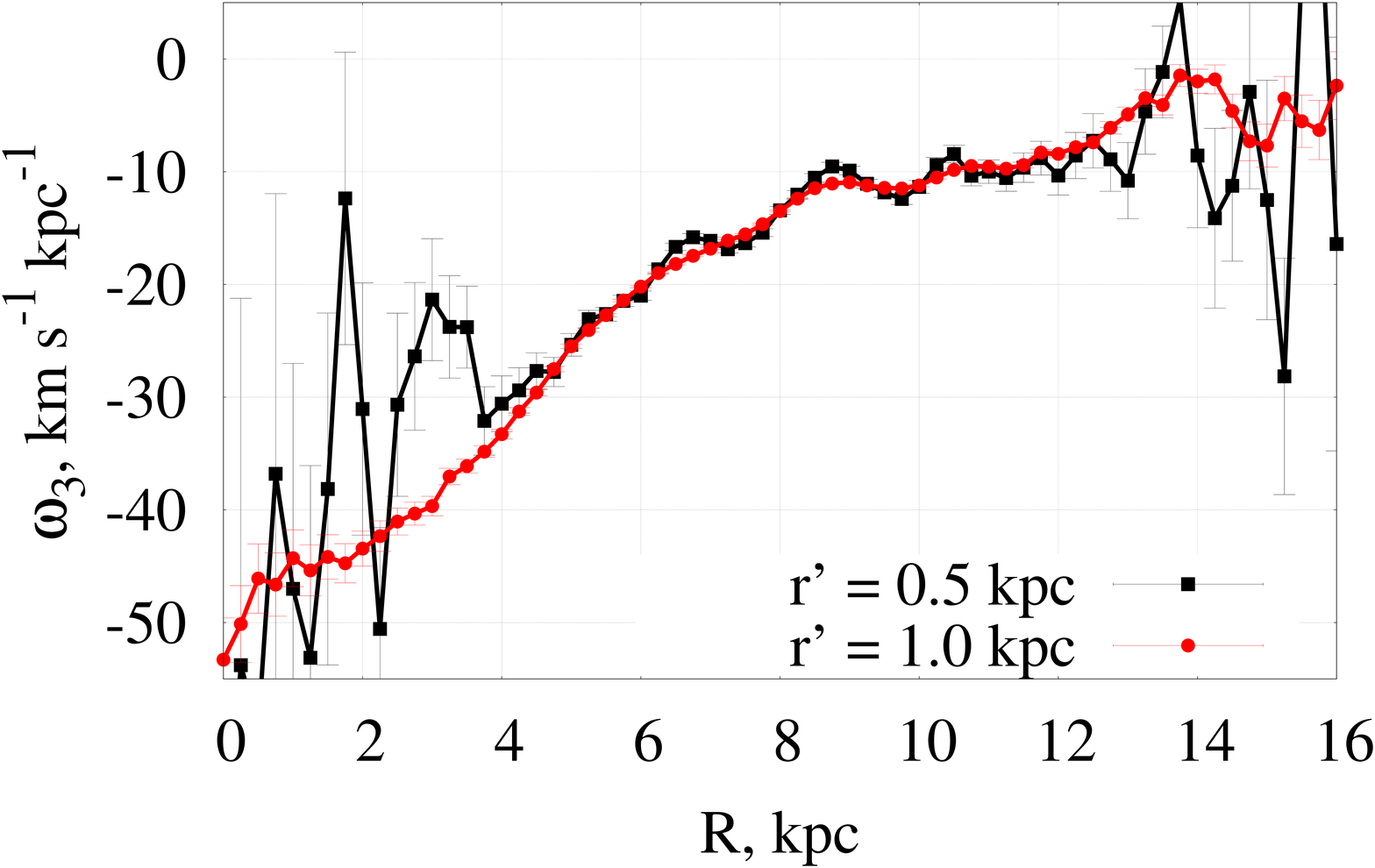}
    \includegraphics{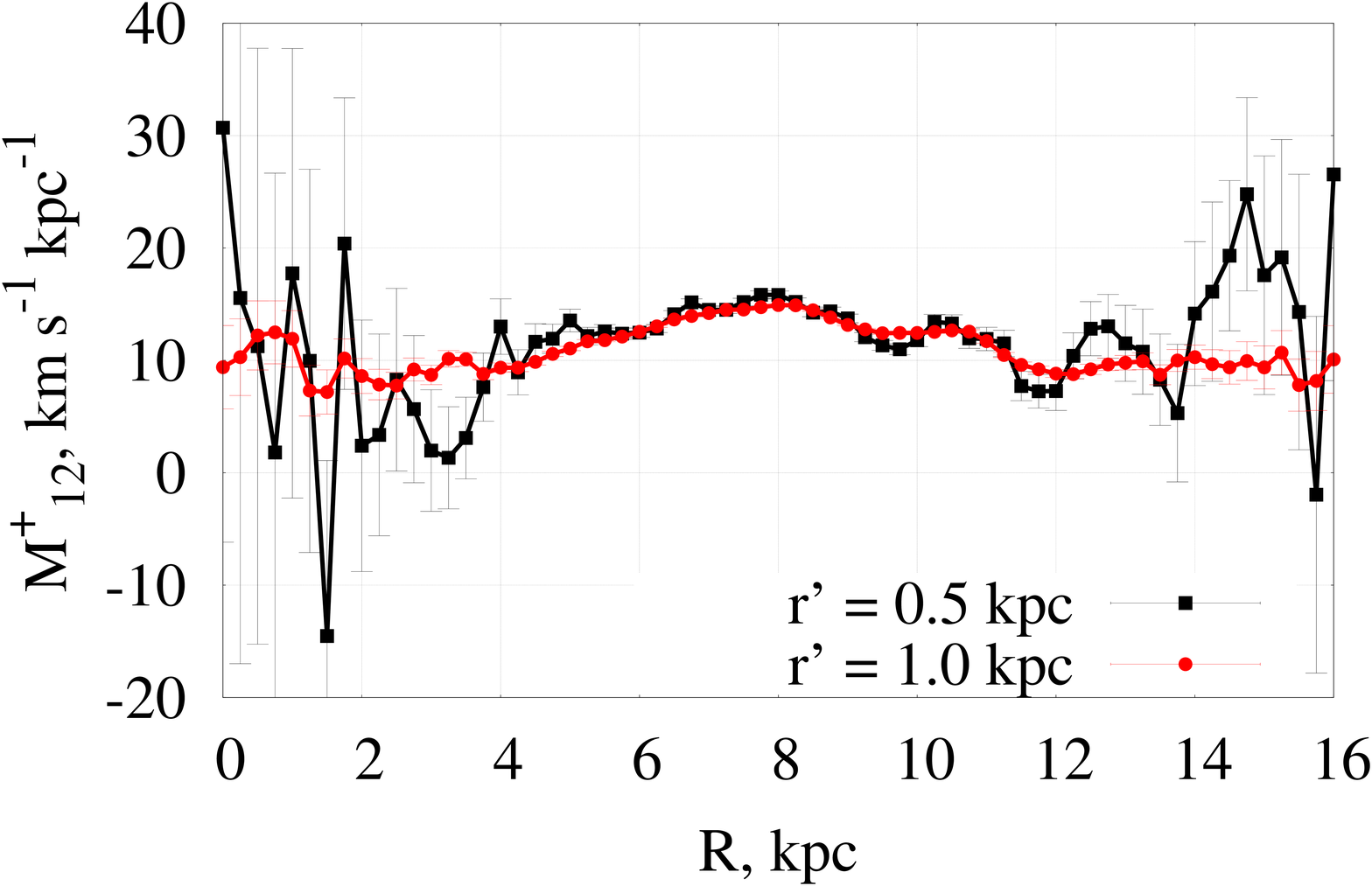}}
\resizebox{\hsize}{!}
   {\includegraphics{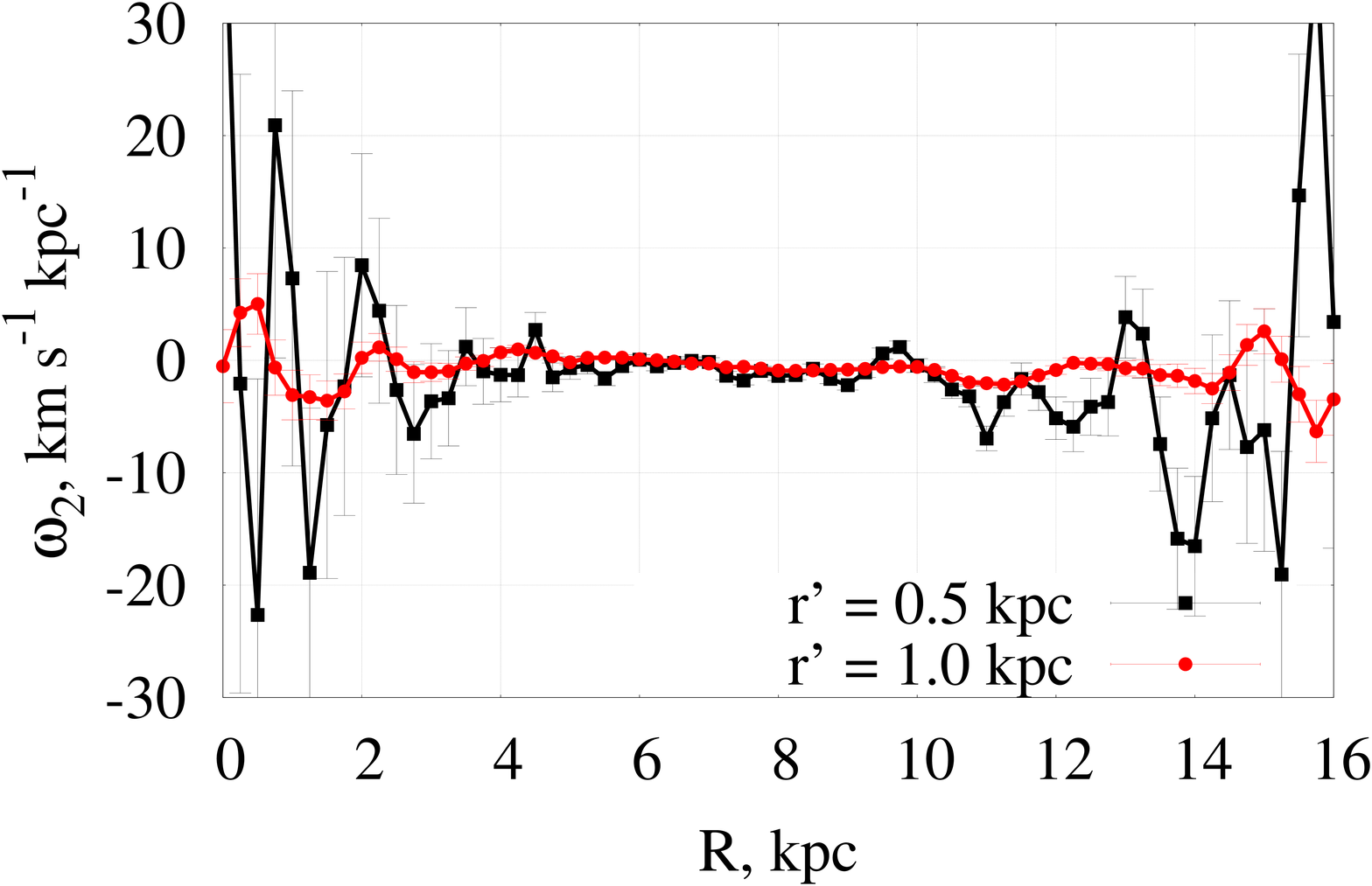}
    \includegraphics{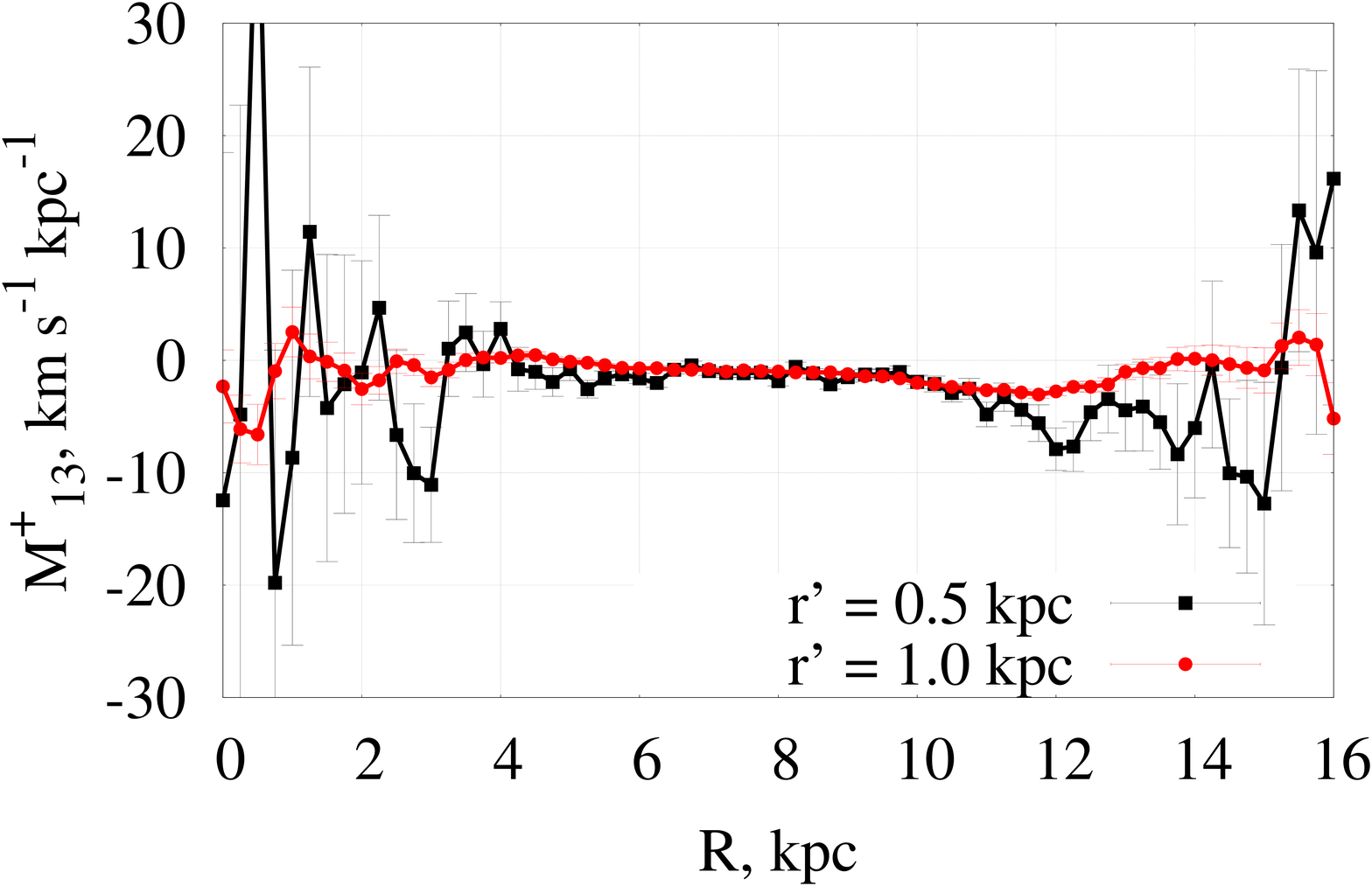}} 
\resizebox{\hsize}{!}
   {\includegraphics{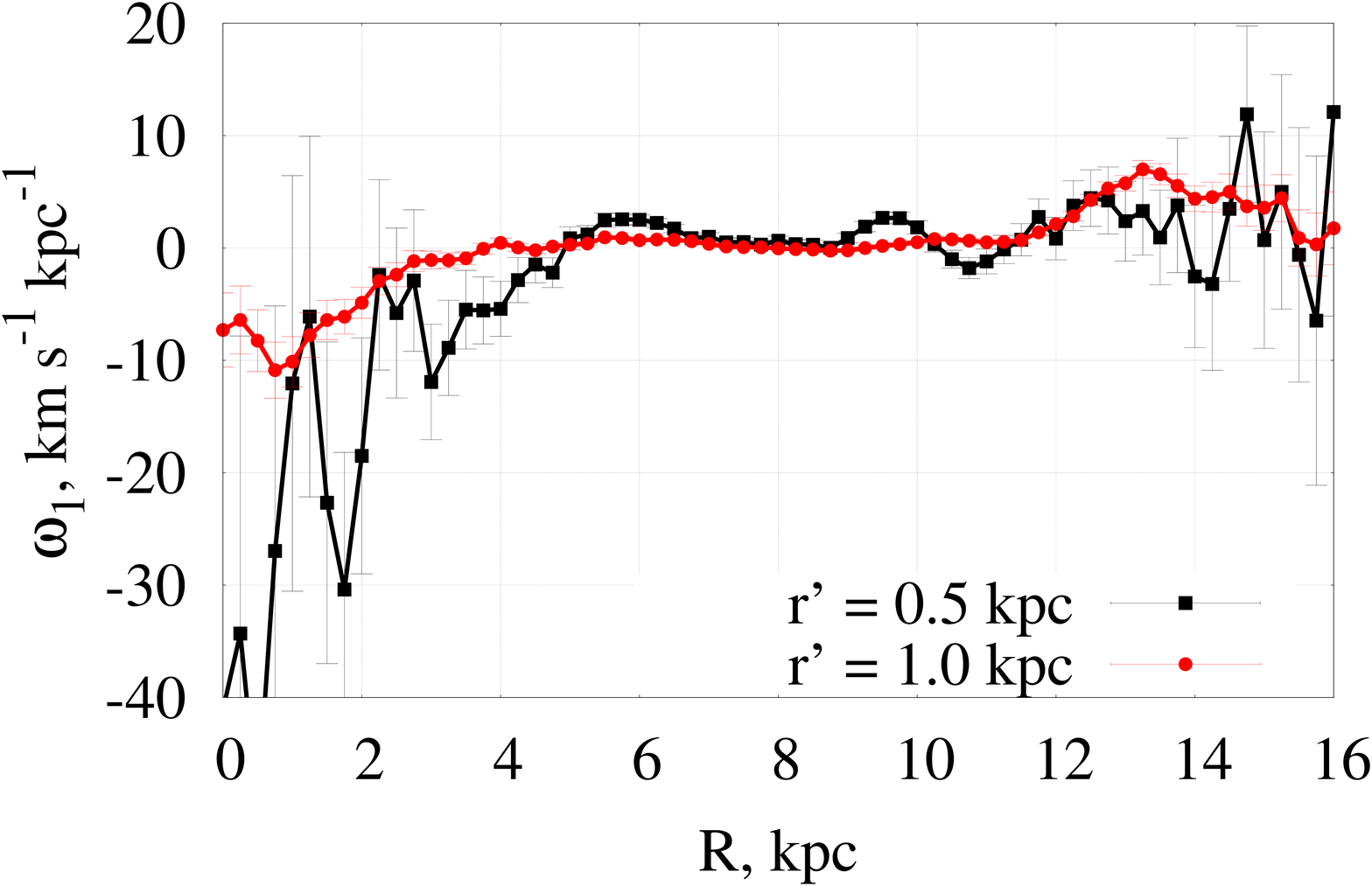}
    \includegraphics{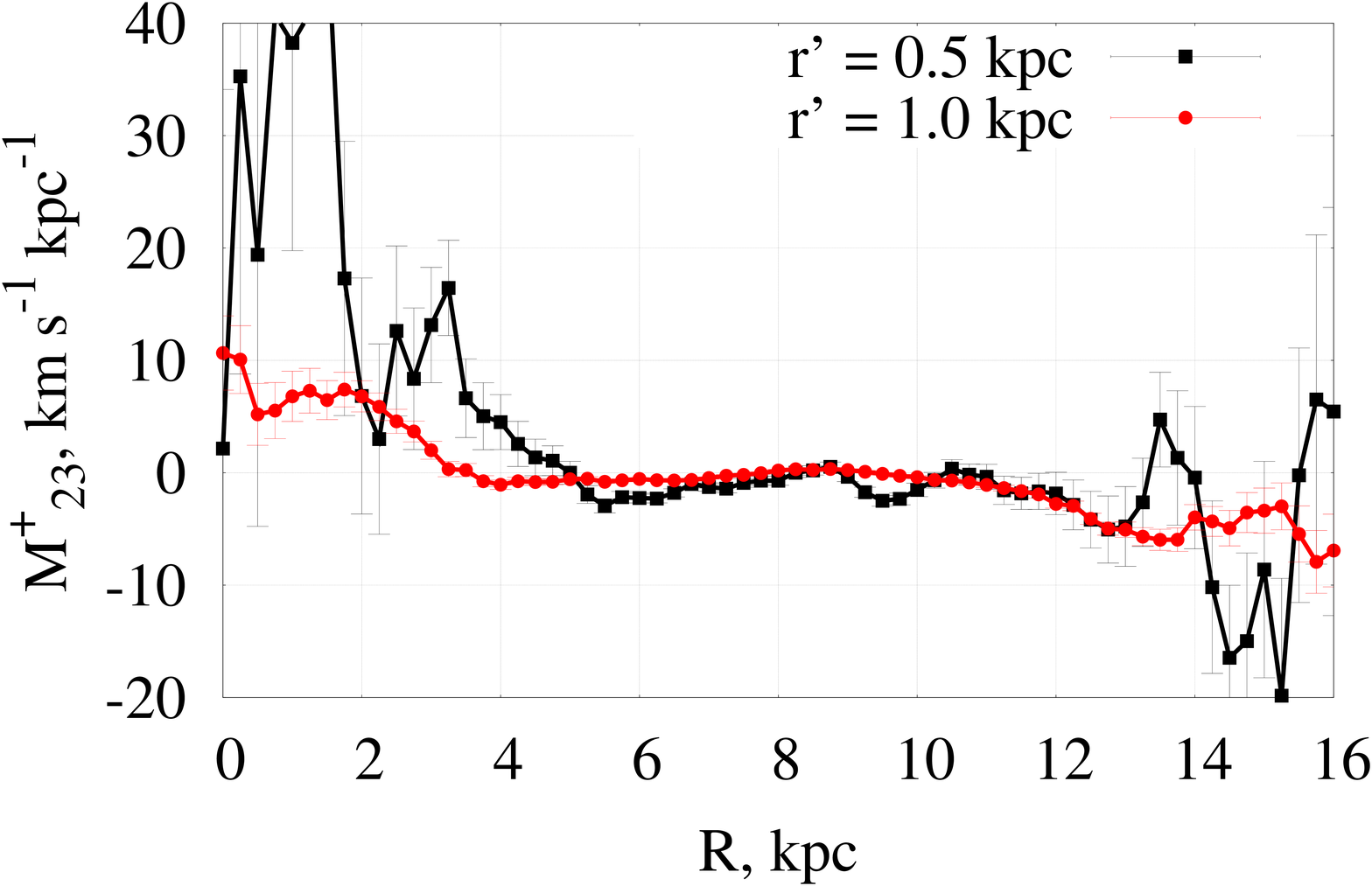}}       
   \caption{The components of local rotation velocities tensor $\omega_3$ (Oort B), $\omega_2, \omega_1 $ (left panel), and the components of deformation velocity tensor $ M^+_{12}$ (Oort A), $ M^+_{13}, M^+_{23}$ (right panel) as a function of the Galactocentric distance.}
\label{fig:OMM_par}%
\end{figure*}

As one can see from Fig.~\ref{fig:OMM_par} (left panel), in the range 4--12~kpc the parameters $\omega_1$ and $\omega_2$ are virtually equal to zero, while $\omega_3$ gradually changes from --30~\kmskpc to --10~\kmskpc and at the Solar distance has the value $\omega_3 = -13.5\pm0.08$~\kmskpc which is found to be in good agreement with those given in numerous papers (for instance, \cite{Vityazev2012, Bovy2017, Vityazev2018, Chengdong2019, Tsvetkov2019, Velichko2020}). The behaviour of the components $\omega_1$, $\omega_2$ and $\omega_3$ indicate that the vectors of instantaneous angular rotational velocity of the corresponding stellar systems are virtually perpendicular to the Galactic plane, and in the modulus almost coincide with the value  $\omega_3$. Out of range 4--12~kpc the components of the rotation vector $\omega_1$ and $\omega_2$ slightly change but we cannot guarantee the behaviour to be realistic, for the reasons mentioned above.

\begin{figure*}
   \centering
\resizebox{\hsize}{!}
   {\includegraphics[width = 610mm]{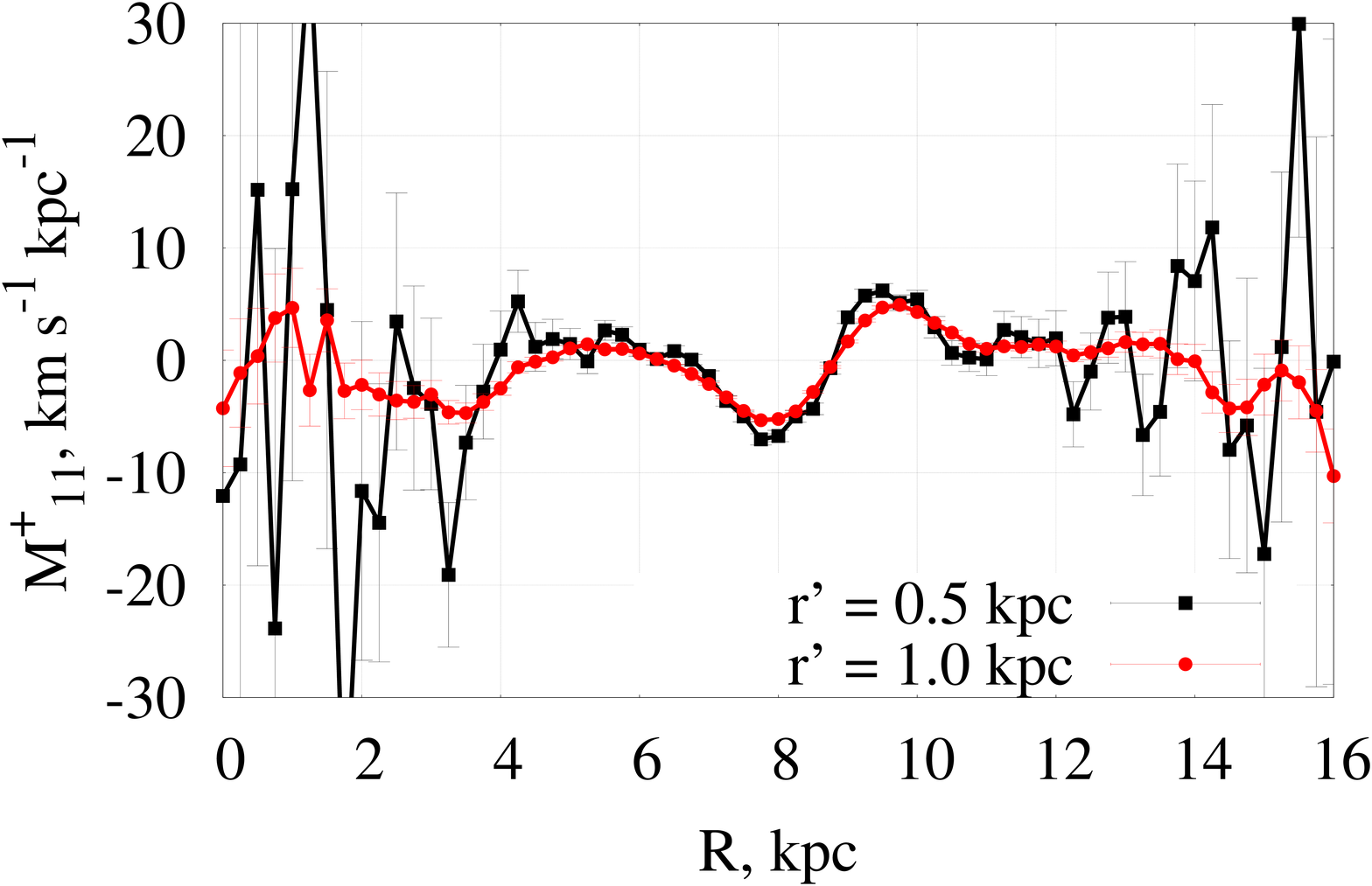}
    \includegraphics{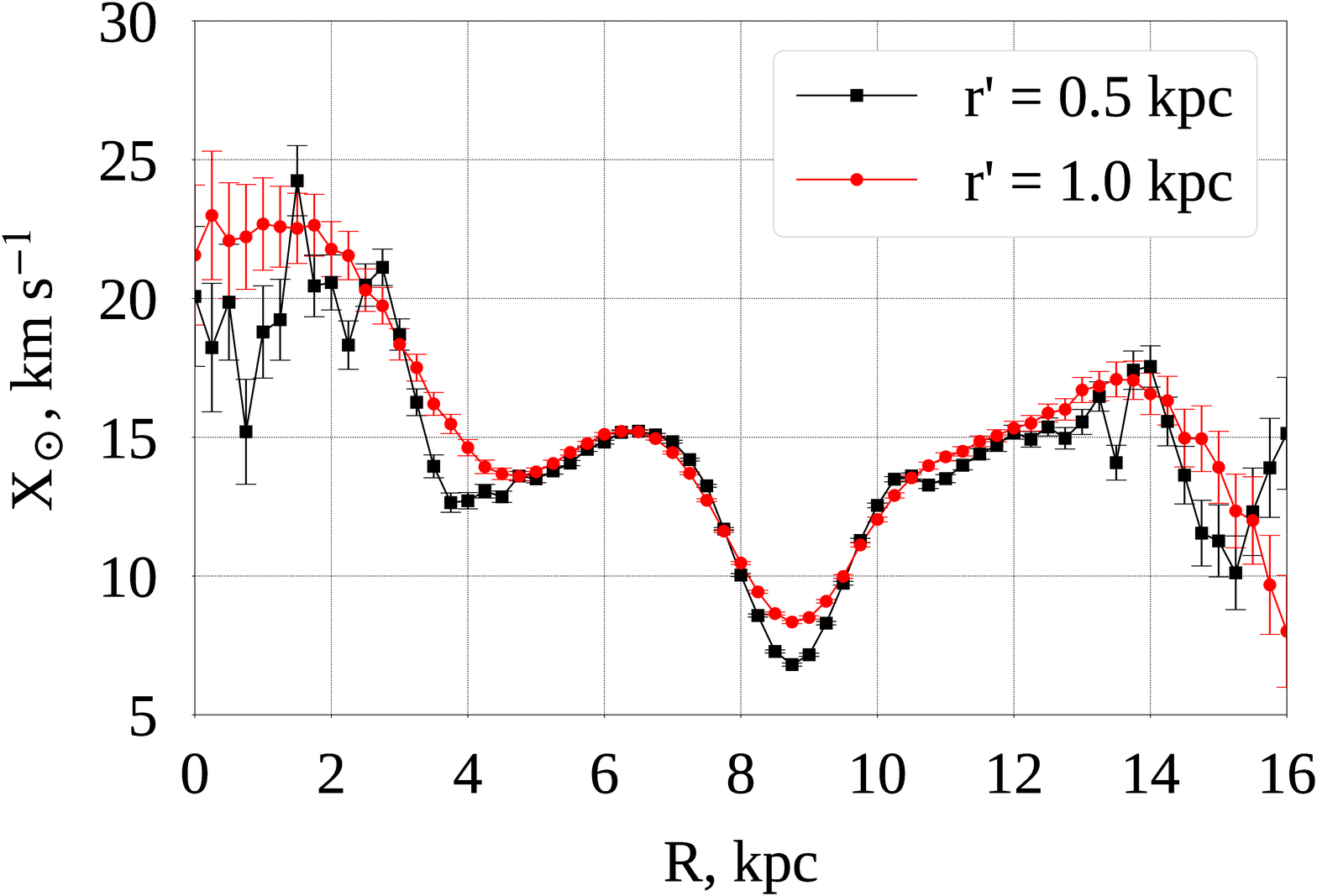}}
\resizebox{\hsize}{!}
   {\includegraphics[width = 610mm]{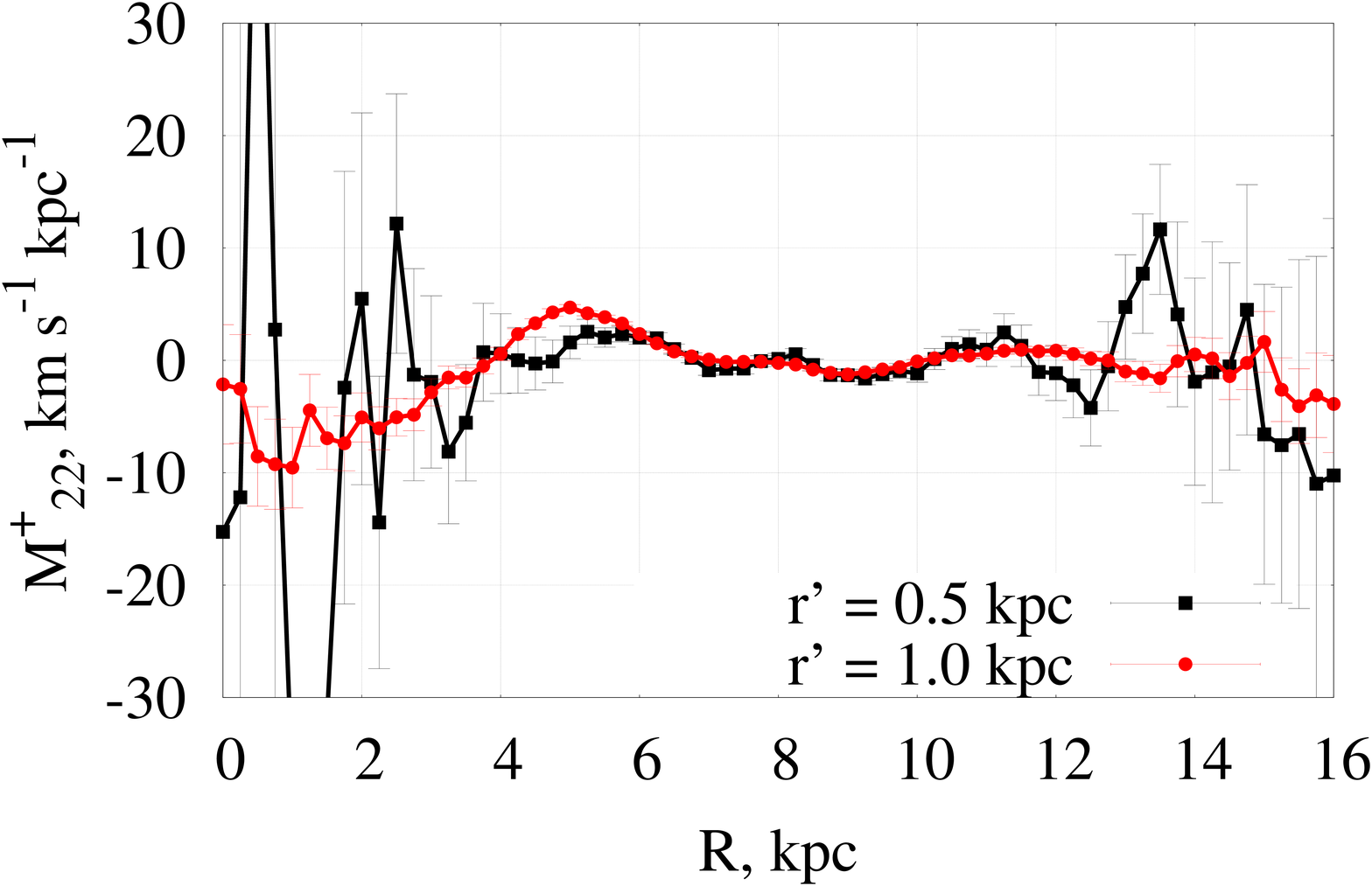}
    \includegraphics{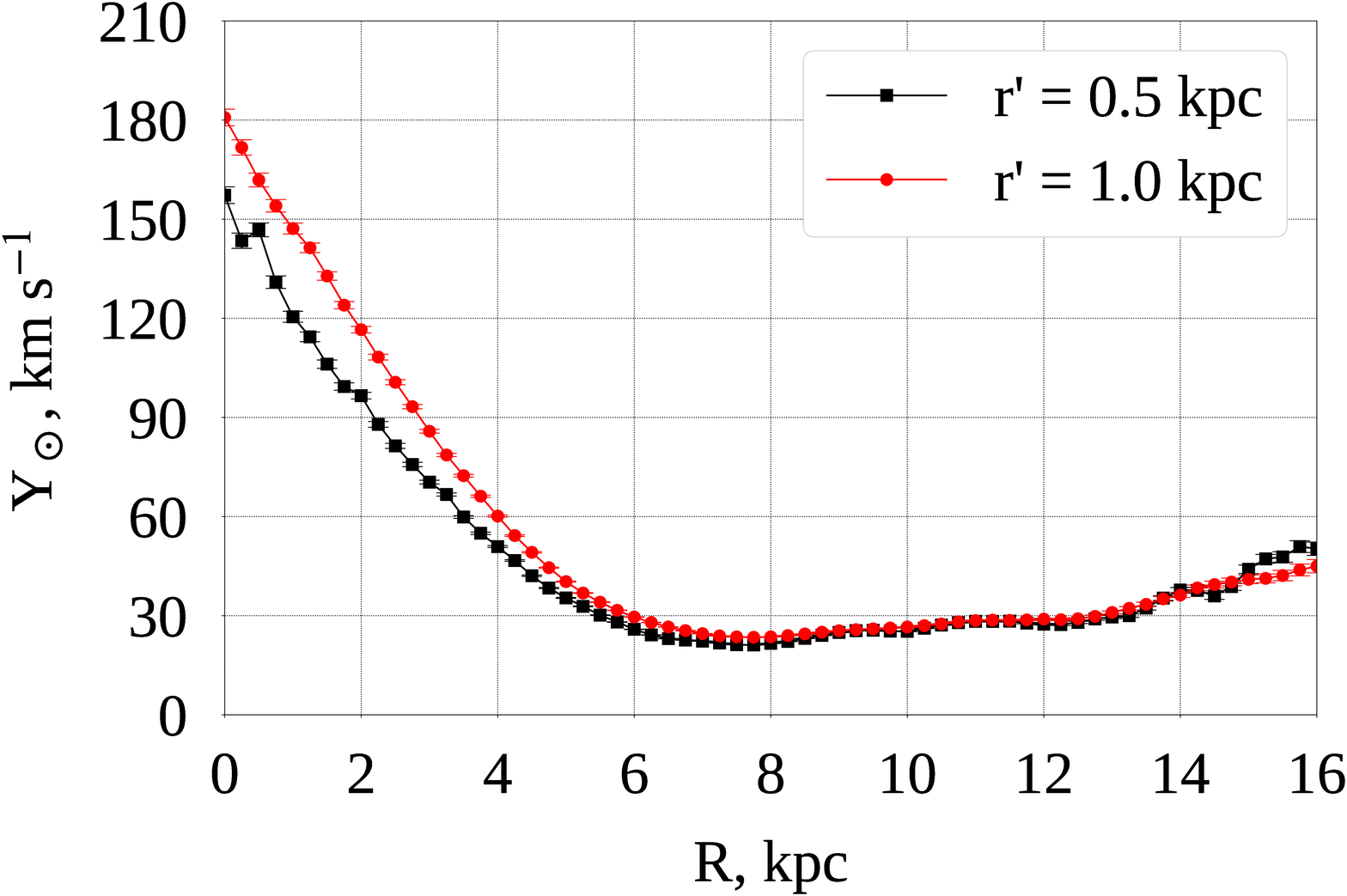}} 
\resizebox{\hsize}{!}
   {\includegraphics[width = 610mm]{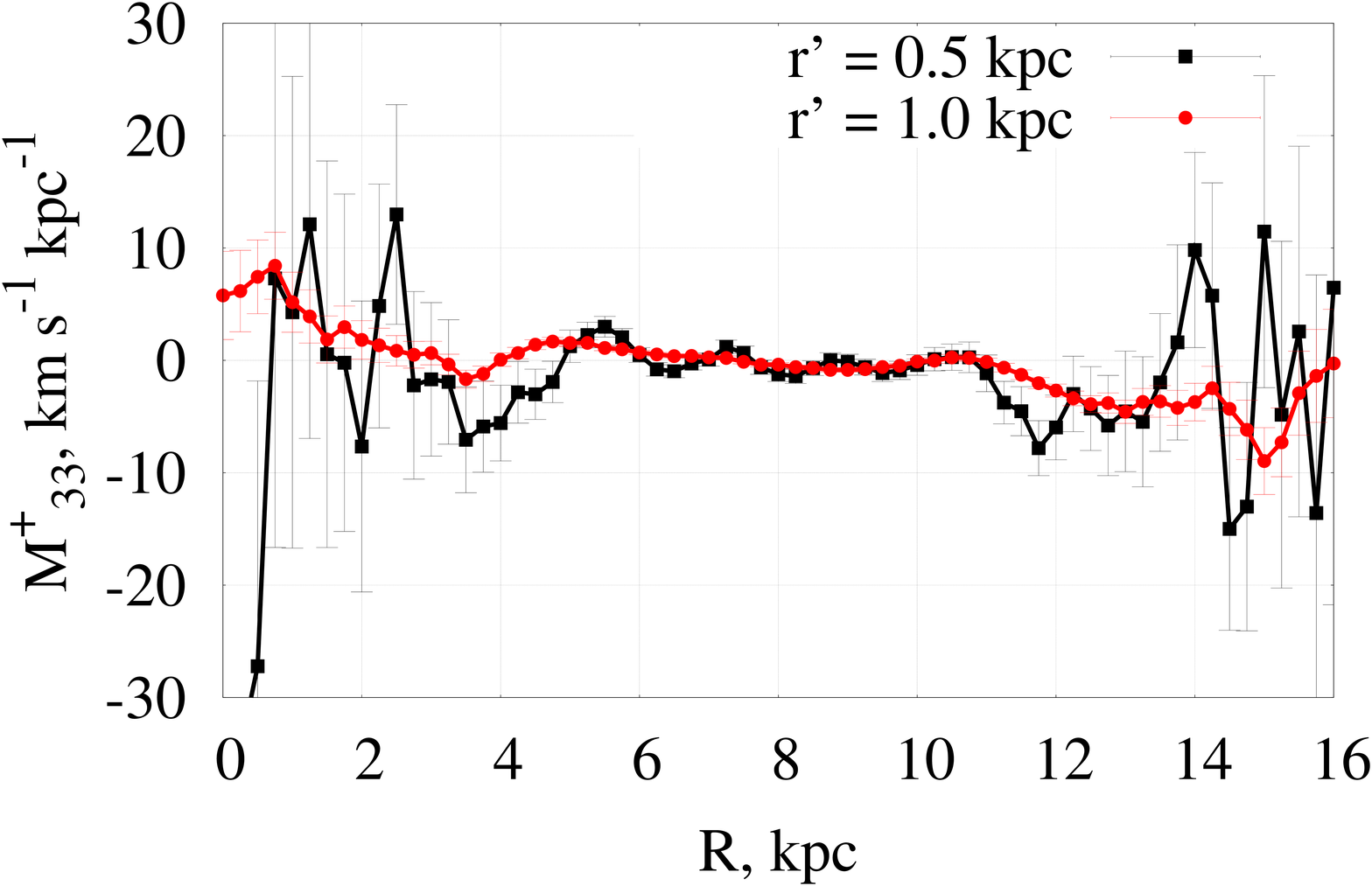}
    \includegraphics{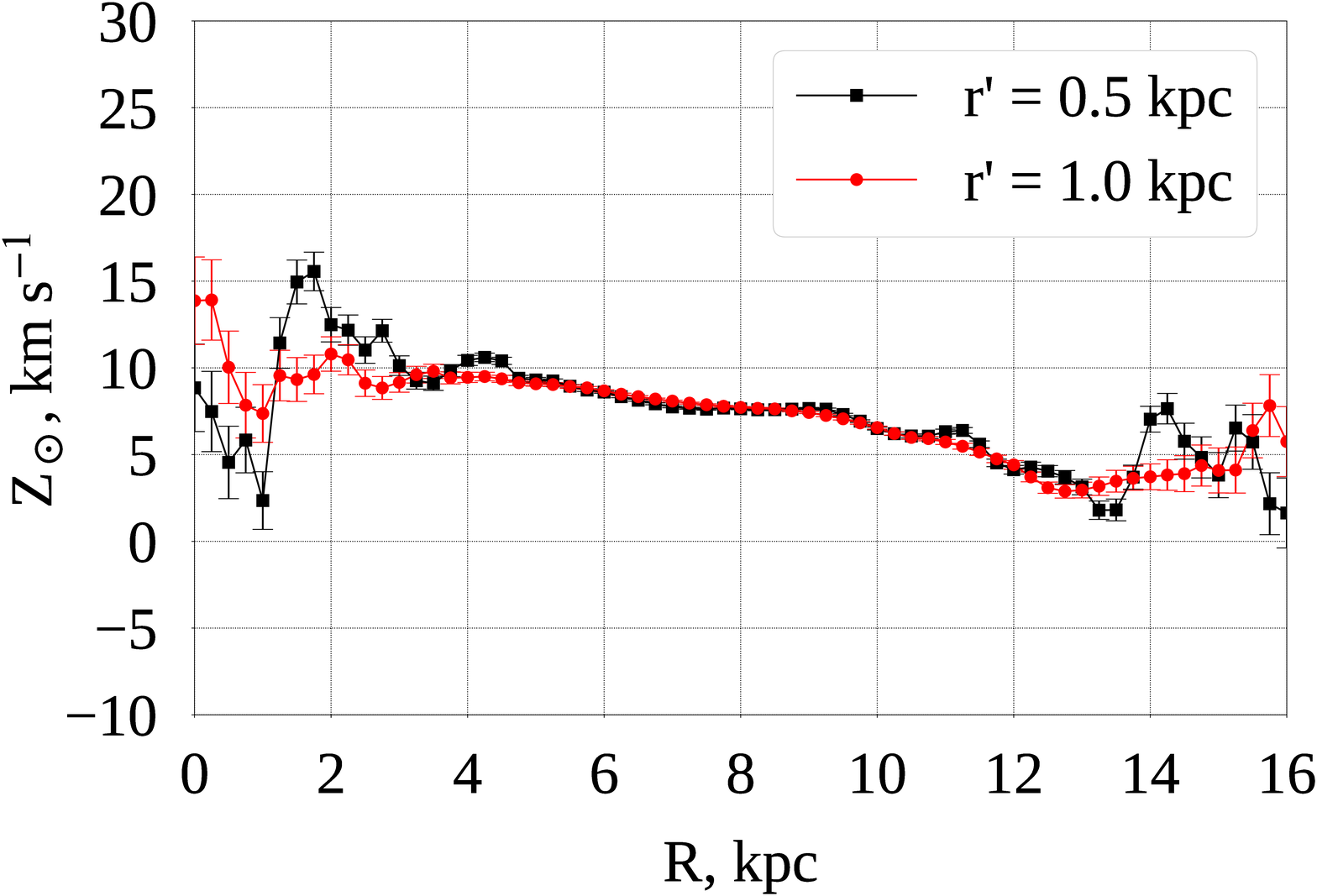}}       
   \caption{The diagonal components of deformation velocity tensor (left panel) and the Solar motion components (right panel) as a function of the Galactocentric distance.}
\label{fig:diagonal_par}%
\end{figure*}

\subsection{Components of the deformation velocity tensor}

The right panel of Fig.~\ref{fig:OMM_par} demonstrates the deformation components $M^+_{12}, M^+_{23}, M^+_{13}$ in the corresponding planes. One can clearly see that in the range 4--8~kpc $M^+_{12}$ gradually increases from 0 to about 16 \kmskpc that followed by a decrease up to 0 at $R =$~12~kpc. The value of $M^+_{12}$ =  14.92$\pm$ 0.08~\kmskpc at the Solar distance also is in good agreement with estimations derived in numerous papers (the same ones as in subsection \ref{sec:omega}). As in the case of rotational components, in the range 4--8~kpc the deformation tensor components $M^+_{13}$ and $M^+_{23}$ are virtually equal to zero, while within the entire distance range 0--16~kpc their behaviour is very similar to that for $\omega_2$ and $\omega_1$ but with opposite sign.

\subsection{Diagonal components of the deformation velocity tensor}

Left panel of Fig.~\ref{fig:diagonal_par} shows dependencies on Galactocentric distance of the kinematic parameters $M^+_{11}$, $M^+_{22}$ and $M^+_{33}$ which characterize contractions/expansions of the stellar systems under study along $x$, $y$, and $z$ axes, respectively. As one can see in left panel of Fig.~\ref{fig:diagonal_par}, the behaviour of $M^+_{11}$ shows pronounced peculiarity in the Solar vicinity. At the same time, in range 4--12 ~kpc $M^+_{22}$ and $M^+_{33}$ do not show such drastic changes in their behavior but demonstrate near zero variations. This indicates relatively small contractions/expansions along the $y$ and $z$ axes within the specified distance range. The $x$-axis direction to the Galactic center (vertex), given inaccurate, may be one of the reasons of the peculiar behaviour of the parameter $M^+_{11}$. However, to unambiguously establish the real reasons of the behaviour requires additional investigations which will be present in our future papers.

\begin{figure}
   \centering
\resizebox{\hsize}{!}
   {\includegraphics{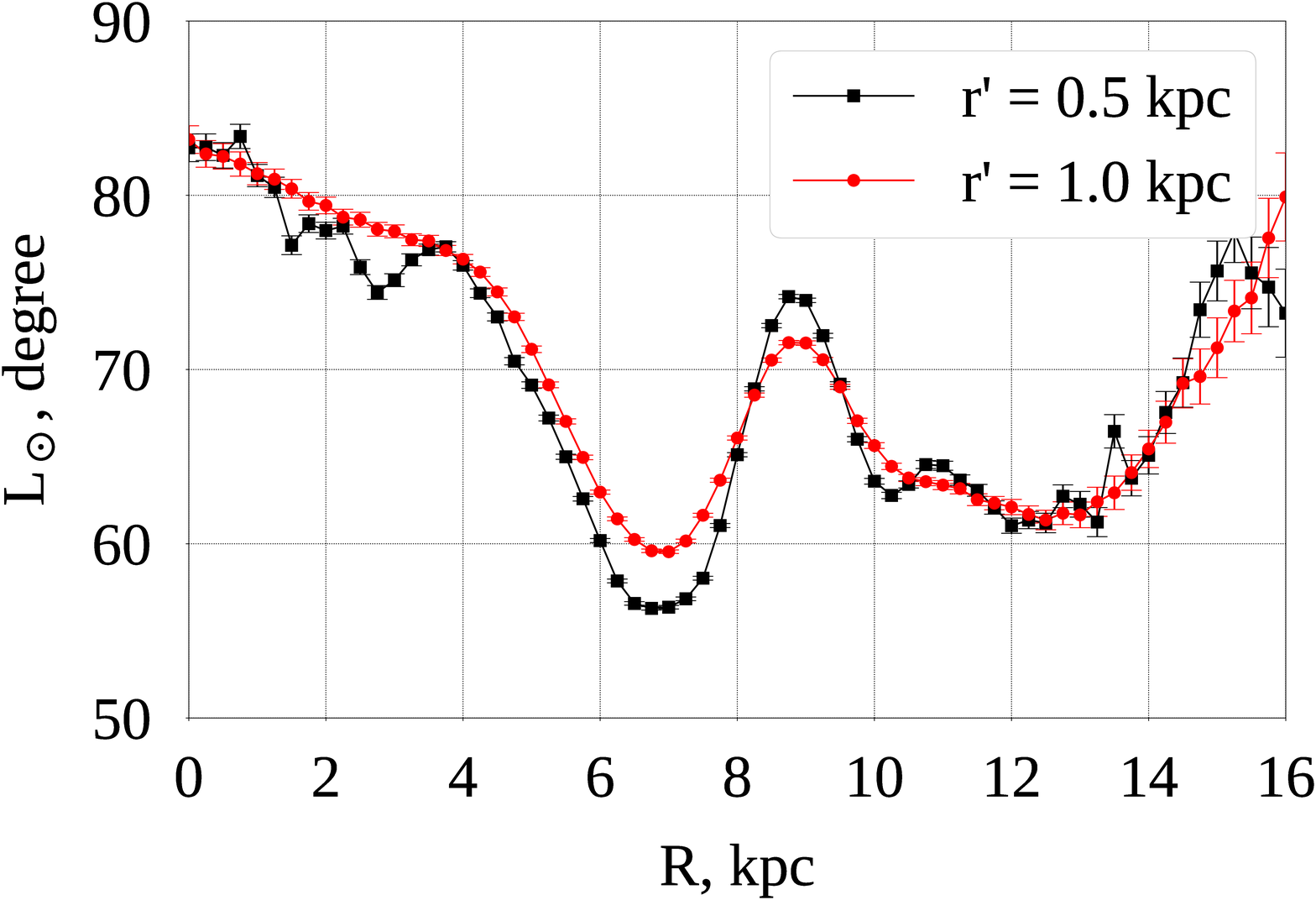}}
\resizebox{\hsize}{!}
   {\includegraphics{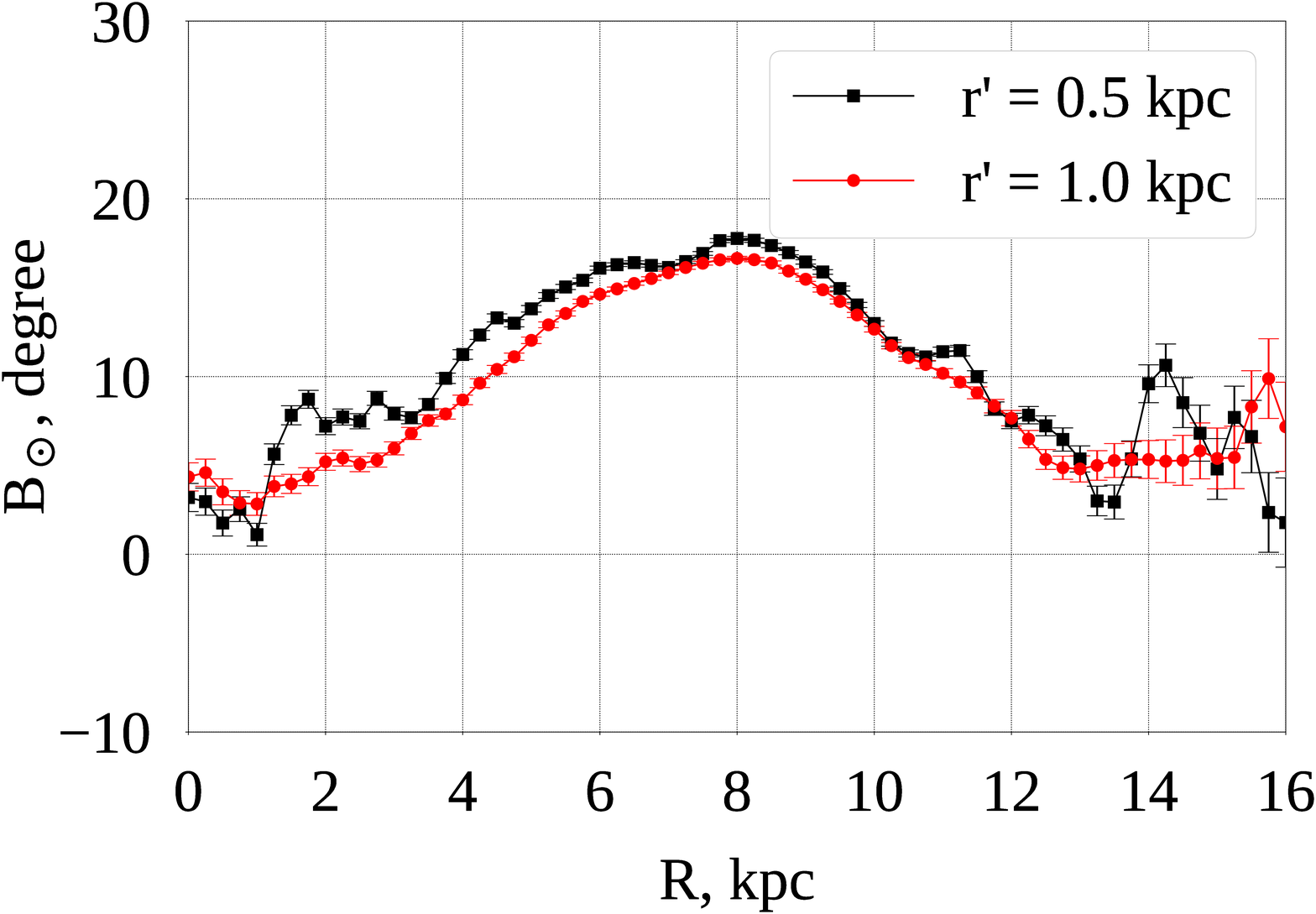}} 
\resizebox{\hsize}{!}
   {\includegraphics{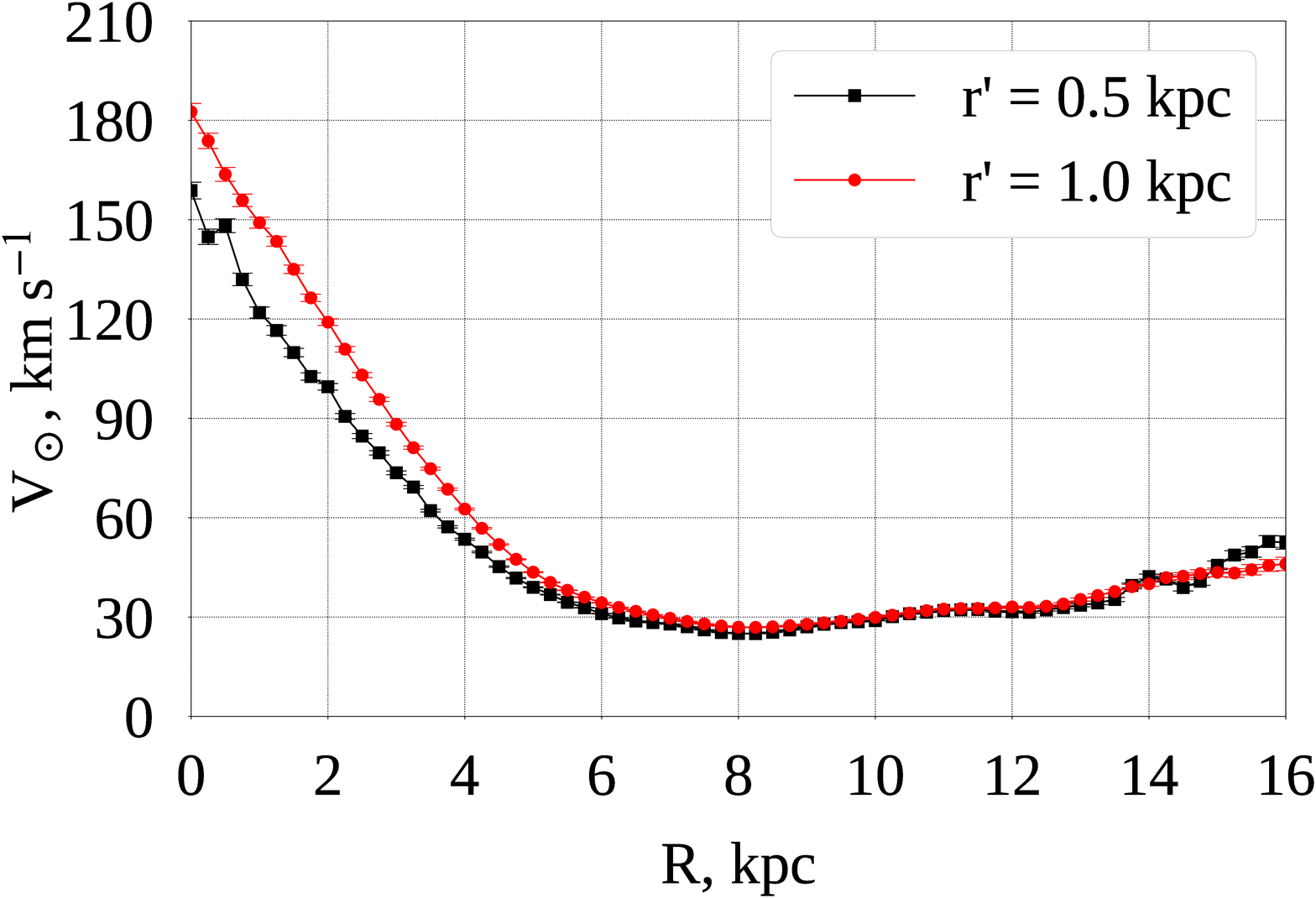}}       
   \caption{Coordinates of the Solar apex and the Solar velocity modulus as a function of the Galactocentric distance.}
\label{fig:sol_motion}%
\end{figure}

\begin{figure}
   \centering
\resizebox{1.1\hsize}{!}
   {\includegraphics{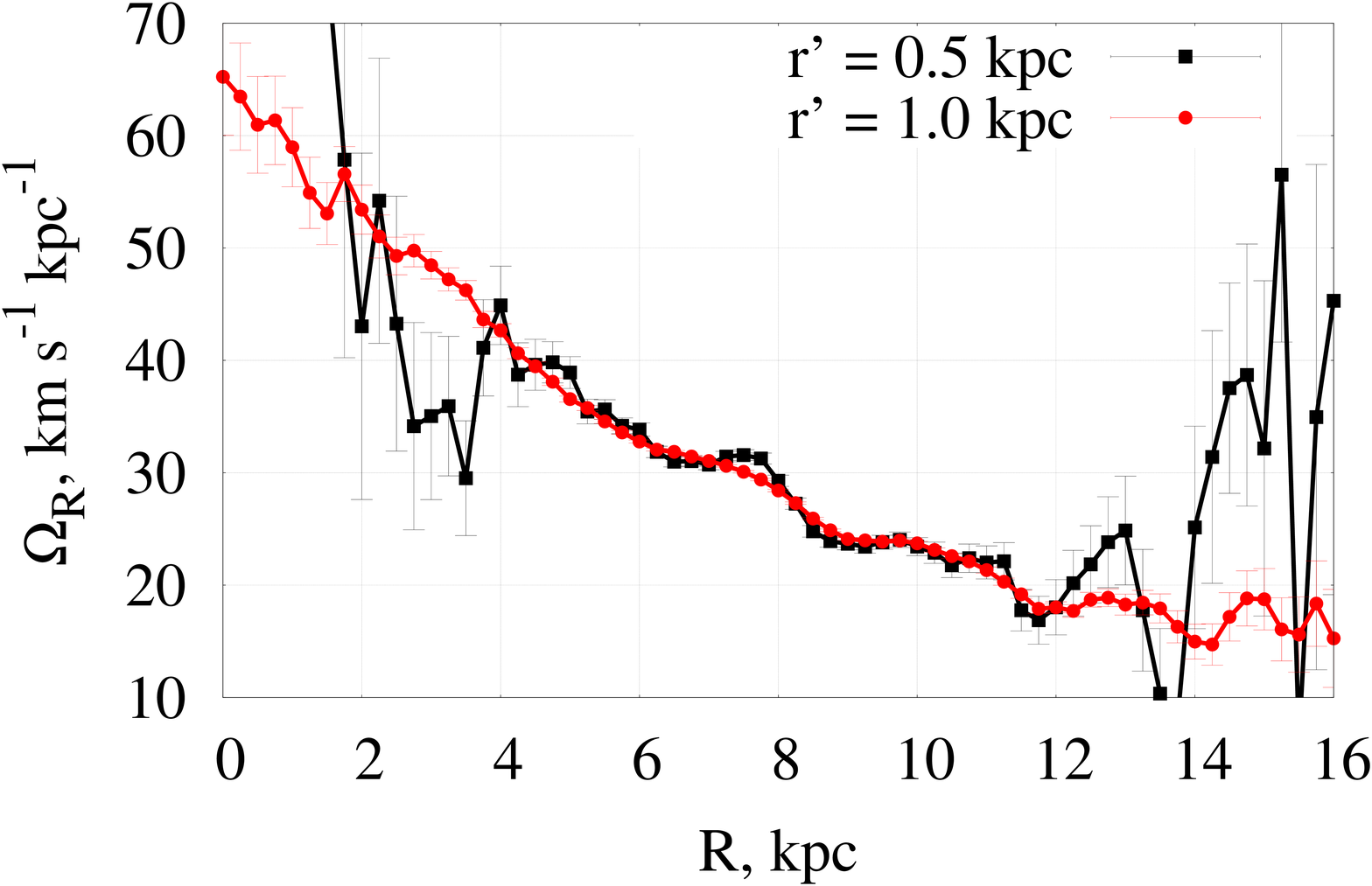}}
  \caption{The modulus of the angular Galactic rotational velocity $|\Omega|$ as a function of the Galactocentric distance.}
\label{fig:ang_vel_mod}%
\end{figure}

\subsection{Components of the Solar velocity relative to different centroids and coordinates of the Solar apex}

Right panel of Fig.~\ref{fig:diagonal_par} shows the components of the linear Solar velocity, while its modulus and
coordinates of the Solar apex are given in Fig,~\ref{fig:sol_motion} as functions of the Galactocentric distance.

As one can see from the right panel of Fig.~\ref{fig:diagonal_par}, the component of the Solar velocity vector $X_\odot$ relative to centroids located within the range from 7 to 10~kpc undergoes noticeable changes, and at $R=8$~kpc has minimal value equal to $8.34\pm0.06$~\kmskpc at $R=8.75$~kpc. It worth noting that within the specified distance range the parameter $M^+_{11}$ also changes significantly. It is obvious that the changes also affect the values of the Solar apex coordinates $L$ and $B$. In this distance range, one can see that there are less noticeable but still obvious variations of the parameters $M^+_{12}$ and $\omega_3$. The values of $X_\odot, Y_\odot$ and $Z_\odot$ relative to the centroid located at the Solar distance, as well as the Solar apex coordinates are in good agreement with those given in other papers (for instance, \cite{Vityazev2014, Velichko2020}), that indicates the behaviour of these Solar velocity components is realistic.

Thus, we have a very interesting result. We know that the Solar velocity relative to the centroids taken with opposite sign is equal to the velocity of centroids relative to the Sun. On the other hand, the Solar velocity modulus relative to the Galactic center is constant. Hence, the behaviour of the centroids' linear velocity components relative to the Galactic center taken with opposite sign has to follow that of the $X_\odot, Y_\odot$ and $Z_\odot$ components. Since the absolute values of the Solar velocity components relative to the Galactic center are not known, it is impossible to establish the offsets on axes for the functions $X_\odot, Y_\odot$ and $Z_\odot$, but their behaviour is determined reliably. In particular, the behaviour of the function $Y_\odot (R)$ have to reproduce that of the Galactic rotational curve $V_{\rm rot}(R)$. Moreover, since the values of the Solar peculiar velocity component $Y_\odot$ have been derived only from the $Gaia$ data, not using Galactocentric distances, we derive the behaviour of the Galactic rotational curve more accurately compared to the traditional estimation present in subsection 5.1.

Irregular behaviour of the component $X_\odot$ may be explained by influence of the Gould Belt's stellar velocity field. The stellar kinematics of this local system with the characteristic radius of about 0.5~kpc is known to differ from that of stars do not belonging to it (see, for instance, \cite{Bobylev2004, Bobylev2014, Alves2020}).

\subsection{Components of the angular Galactic rotational velocity}

\begin{figure*}
   \centering
\resizebox{\hsize}{!}
   {\includegraphics{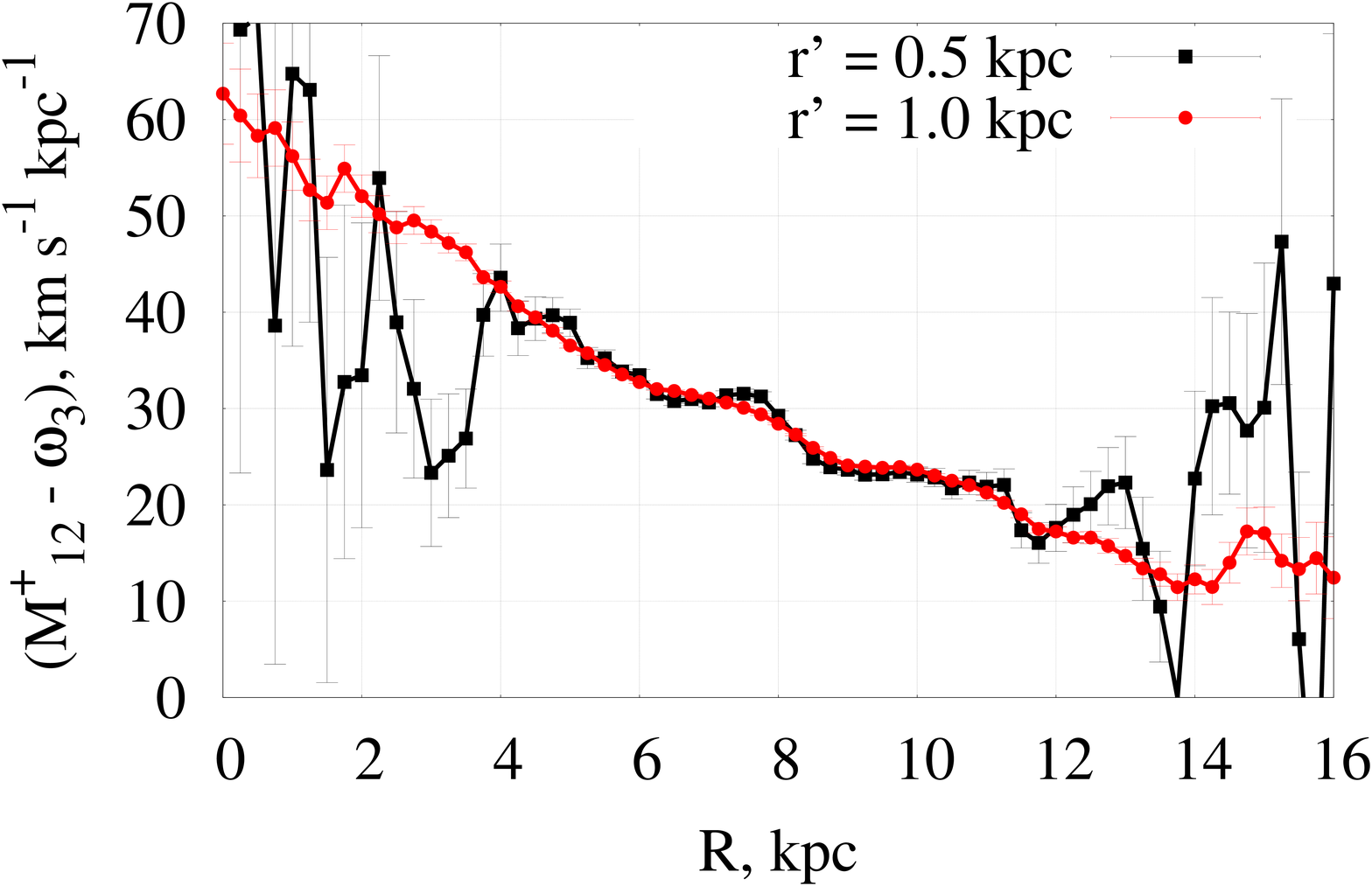}
    \includegraphics{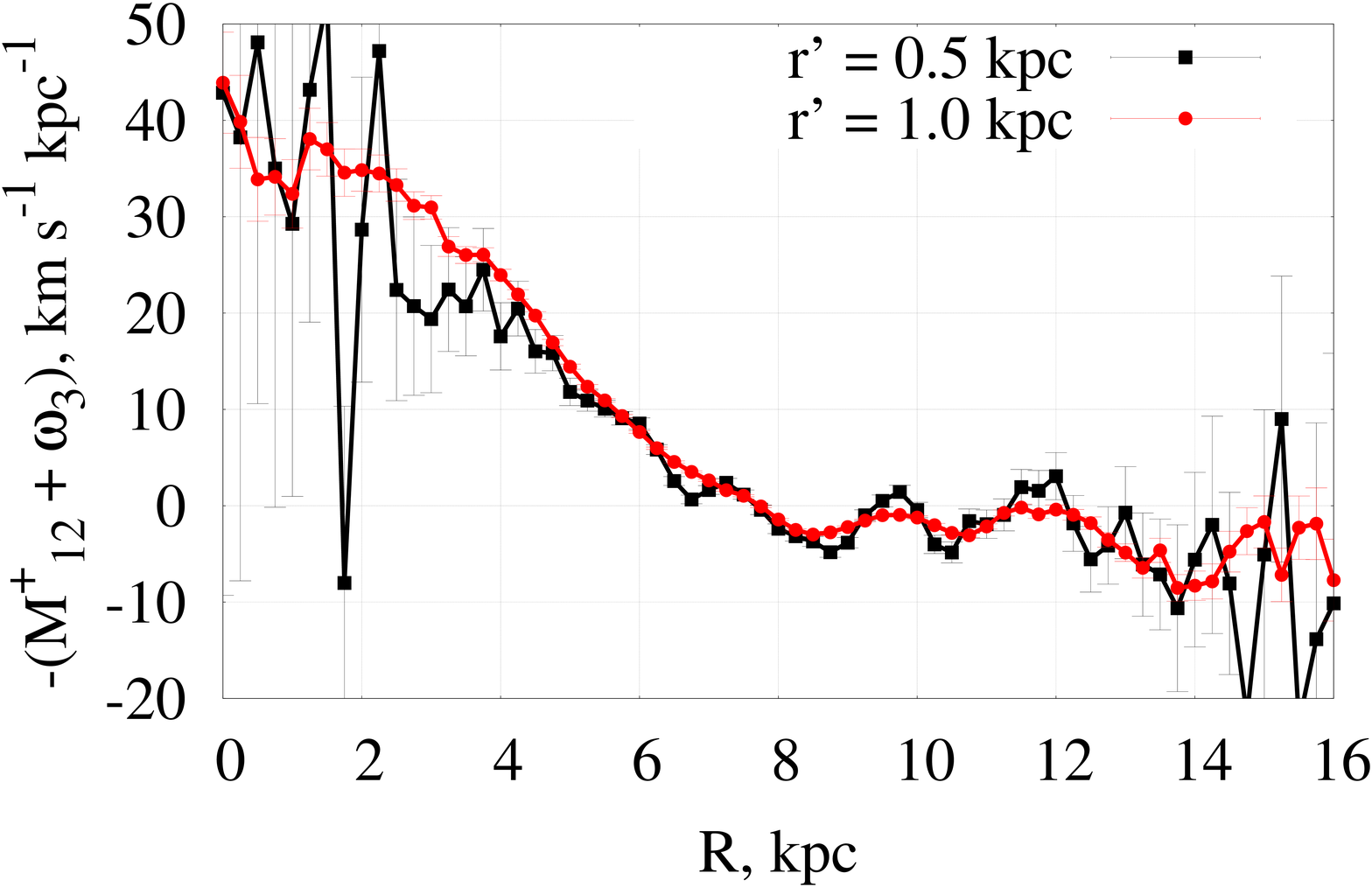}}
\resizebox{\hsize}{!}
   {\includegraphics{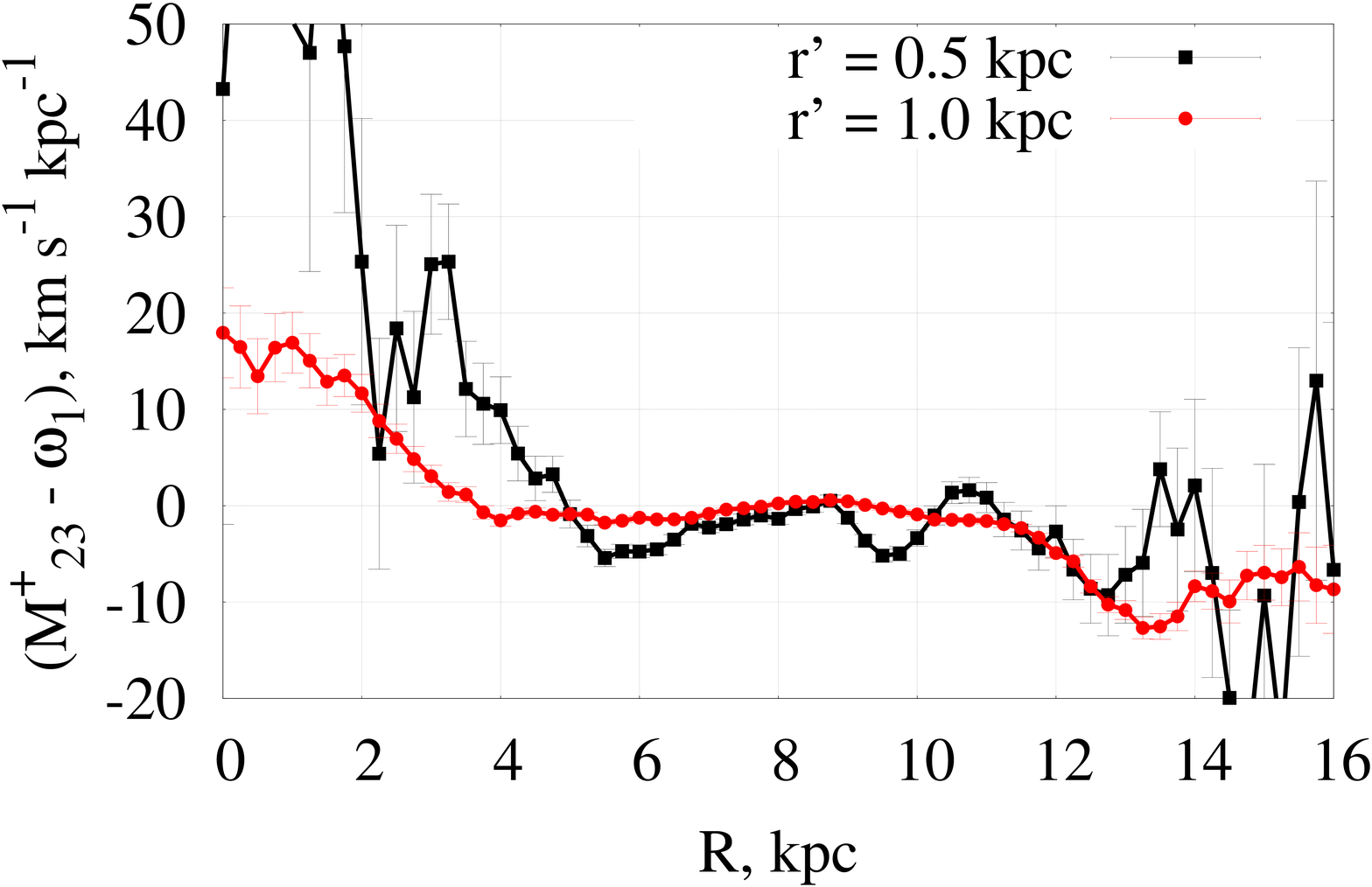}
    \includegraphics{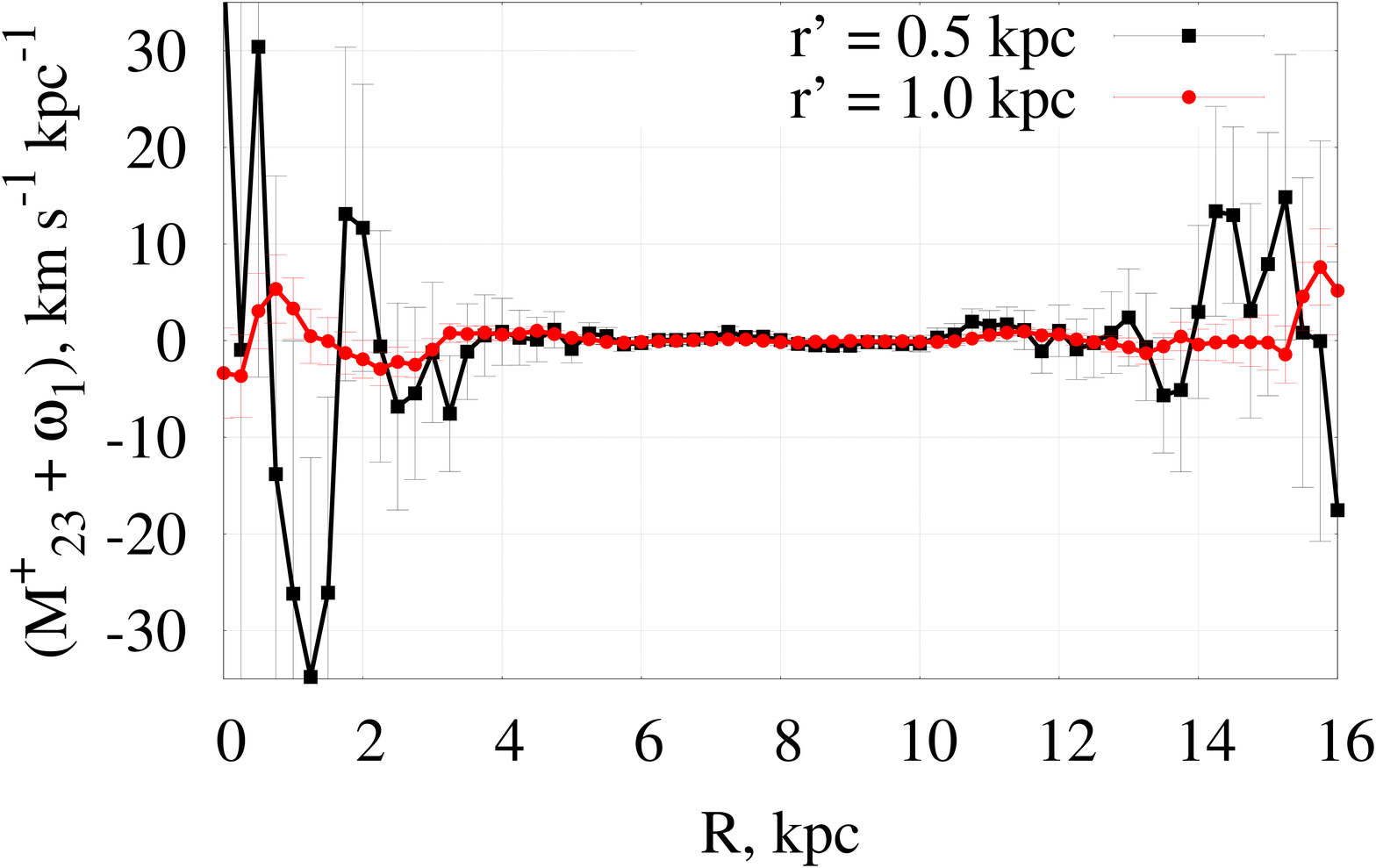}} 
\resizebox{\hsize}{!}
   {\includegraphics{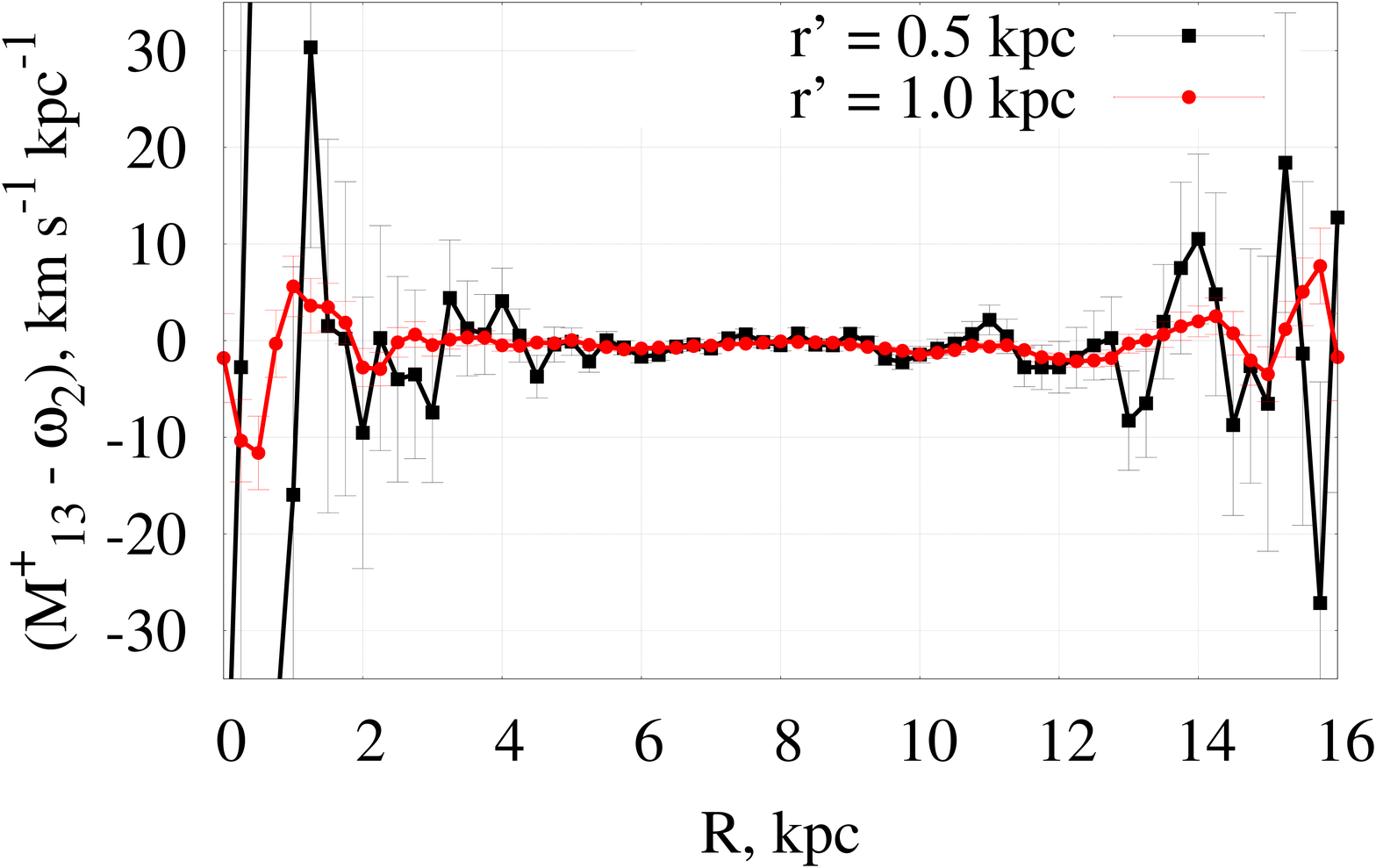}
    \includegraphics{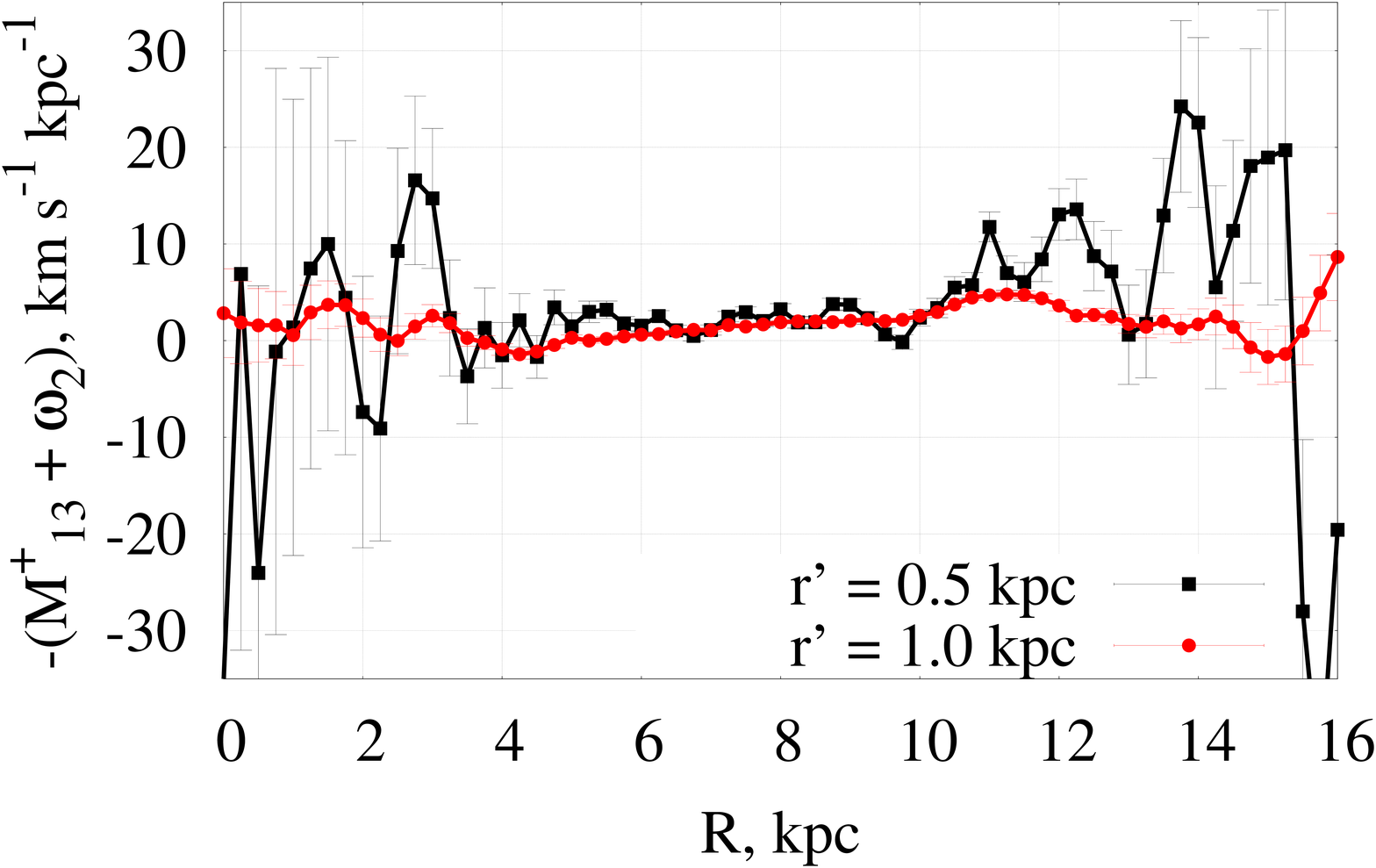}}       
   \caption{The component of the angular Galactic rotational velocity vector $\bf \Omega$ and gradients of components Galactocentric velocity as a function of the Galactocentric distance.}
\label{fig:ang_vel}
\end{figure*}

Fig.~\ref{fig:ang_vel} shows the components of the angular Galactic rotational velocity vector $\bf \Omega$ derived from the analysis of the stellar systems under examination. Fig.~\ref{fig:ang_vel_mod} shows the vector's modulus as a function of the Galactocentric distance $R$. One can clearly see from the figures that in the range 4--12~kpc the components of the angular Galactic rotational velocity are nearly zero, and therefore the Galactic rotation vector within the range are virtually perpendicular to the Galactic plane. The value of $\Omega$ at the Solar distance is about 30~\kmskpc.

Finally, We emphasize that the kinematic parameters derived in this work for the Solar centroid are very close to those published in earlier works mentioned above. This indicates that all the values derived for other centroids located along the $x$-axis are adequate. 

\section{Interpretation of the kinematic parameters given in the cylindrical coordinate system}

To interpret the derived results and get some information about the Galaxy as a whole we move to the Galactocentric coordinate system (\cite{Miyamoto1993, Bobylev2006, Vityazev2012}, Fig.\ref{fig:GCS}). As before, here we have in mind that the components of the rotational and deformation velocity tensors defined in the vertical planes $xz$ and $yz$ have opposite signs in the northern and southern Galactic hemispheres. When solving the equations by the LSM this regularity may be broken due to stellar distribution over the sphere is not uniform, especially in spatial areas which are distant from the Sun. This may result in biases in the least squares solutions.
 
The elements of the rotational and deformation tensors in the local Galactic coordinate system are related to the stellar velocity field components $V_R$, $V_\theta$ and $V_z$ in the cylindrical Galactocentric coordinate system ($R, \theta, z$) in the following way:

\begin{figure}
   \centering
\resizebox{\hsize}{!}
   {\includegraphics[width = 710mm]{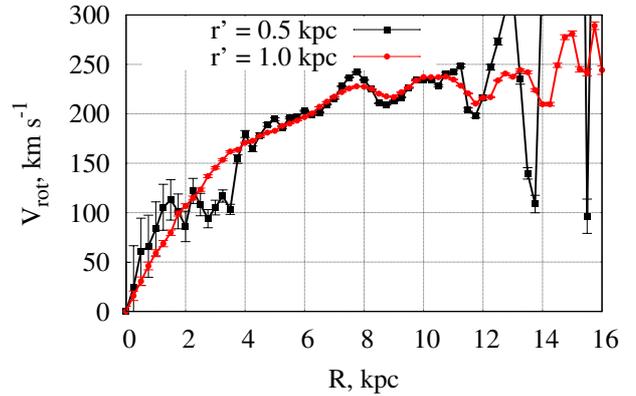}}
  \caption{Circular velocity of Galaxy as a function of the Galactocentric distance ( Galactic rotation curves).}
\label{fig:Vrot}
\end{figure}

\begin{figure}
   \centering
\resizebox{\hsize}{!}
   {\includegraphics{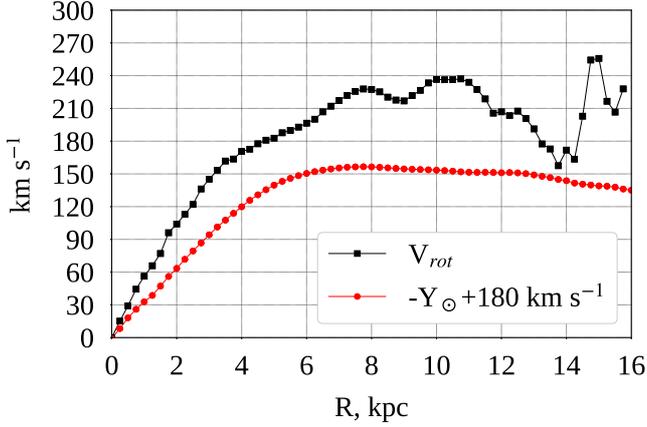}}
  \caption{Comparison of the behaviours of the Galactic rotational curves derived from the equation \ref{eq:Vrot} and using the component $Y_\odot$.}
\label{fig:Vrot-Y}
\end{figure}

\begin{gather}
M = 0.5
\left( \begin{matrix}
  2\,M^+_{11}         & M^+_{12} + M^-_{12} & M^+_{13} + M^-_{13} \\[6pt]
  M^+_{21} + M^-_{21} & 2\,M^+_{22}            & M^+_{23} + M^-_{23} \\[6pt]
  M^+_{31} + M^-_{31} & M^+_{32} + M^-_{32} & 2\,M^+_{33} 
  \end{matrix} \right)\nonumber
  =
  \end{gather}
  
  \begin{gather}
  =
\left( \begin{matrix}
  \frac{\partial V_R}{\partial R} & \frac{1}{R}\frac{\partial V_R}{\partial \theta} - \frac{V_\theta}{R} & \frac{\partial V_R}{\partial z} \\[10pt]
   \frac{\partial V_\theta}{\partial R}& \frac{1}{R}\frac{\partial V_\theta}{\partial \theta} + \frac{V_R}{R} & \frac{\partial V_\theta}{\partial z} \\[10pt]
   \frac{\partial V_z}{\partial R}&\frac{1}{R}\frac{\partial V_z}{\partial \theta} &\frac{\partial V_z}{\partial z} 
  \end{matrix} \right) 
\end{gather}

\subsection {Galactic rotational velocity in the centroids' vicinity}

 The components of the Galactocentric velocities $V_R$, $V_\theta$ and $V_z$ as well as their derivatives can be expressed via the combinations of the O--M model parameters. Thus,

\begin{equation}
\frac{V_\theta}{R} + \frac{1}{R} \frac{\partial V_R} {\partial \theta} = (\omega_3 - M^+_{12}).
\end{equation}

Since the stellar system's angular rotational velocity components in the $xz$ and $yz$ planes (Fig.~\ref{fig:OMM_par}) are nearly 
zero within the distance range 4--12~kpc, the radical expression \ref{eq:Omega_R} contains only the first term. As a result, 
the stellar velocity field within the specified distance range may be considered as axisymmetric. Therefore, we assume that 
\[
\frac{\partial V_R}{\partial \theta} = 0. 
\]
Then, given the Galactic rotation is opposite to the direction of reference of the azimuth angle $\theta$, we can compute the circular velocity of the local coordinate system $V_{\rm rot}$ (i.e. Galactic rotational velocity) as follows:
\begin{equation}
V_{\rm rot} = -V_\theta = (M^+_{12} - \omega_3)_R R = \Omega_R R
\label{eq:Vrot}
\end{equation}

Fig.~\ref{fig:Vrot} shows the linear circular Galactic rotational velocity $V_{\rm rot} = \Omega_{R} R$ as a function of the Galactocentric distance $R$. As one can see from the figure, the values of $V_{\rm rot}$ determined in the Solar vicinity is in good agreement with those given in numerous papers (see above), and it is equal to about 227.36$\pm$0.11~\kms. The dependency, commonly referred to as the Galactic rotational curve, is reliable only within the distance range 4--12~kpc. Unfortunately, out of the specified range the values of $V_{\rm rot}$ should not be trusted. Nevertheless, as mentioned above, Fig.~\ref{fig:Vrot} demonstrates the entire distance range from 0 to 16~kpc to get insight into values and behaviour of $V_{\rm rot}$ near the Galactic center and at its outer part.

For comparison, Fig.~\ref{fig:Vrot-Y} also shows the dependency of $Y_\odot$ on $R$, derived only from the peculiar Solar velocity relative to the centroids given with opposite sign. These two curves have been align by ordinate at $R=$~0~kpc.

\subsection {Radial gradient of the Galactic rotational velocity}

The radial gradient of the Galactic rotational velocity, or the slope of the Galactic rotational curve, is defined by the relation:

\begin{equation}
\frac{\partial V_{\rm rot}}{\partial R} = -(M^+_{12} + \omega_3)
\label{eq:Vrot_slope}
\end{equation}

The top panel of Fig.\ref{fig:ang_vel} shows $\partial V_{\rm rot}/\partial R$ as a function of $R$. It is clearly seen that the values and behaviour of the slope within the distance range from 0 to about 8~kpc are noticeably different from those in the range 8--16~kpc. In the latter case, the curve's slope decreases, and its magnitude oscillates around $\partial V_{\rm rot}/\partial R$= --2 - --5~\kmskpc, that, in principle, does not contradict to the behaviour of the Galactic rotational curve.

\begin{figure}
   \centering
\resizebox{\hsize}{!}
   {\includegraphics{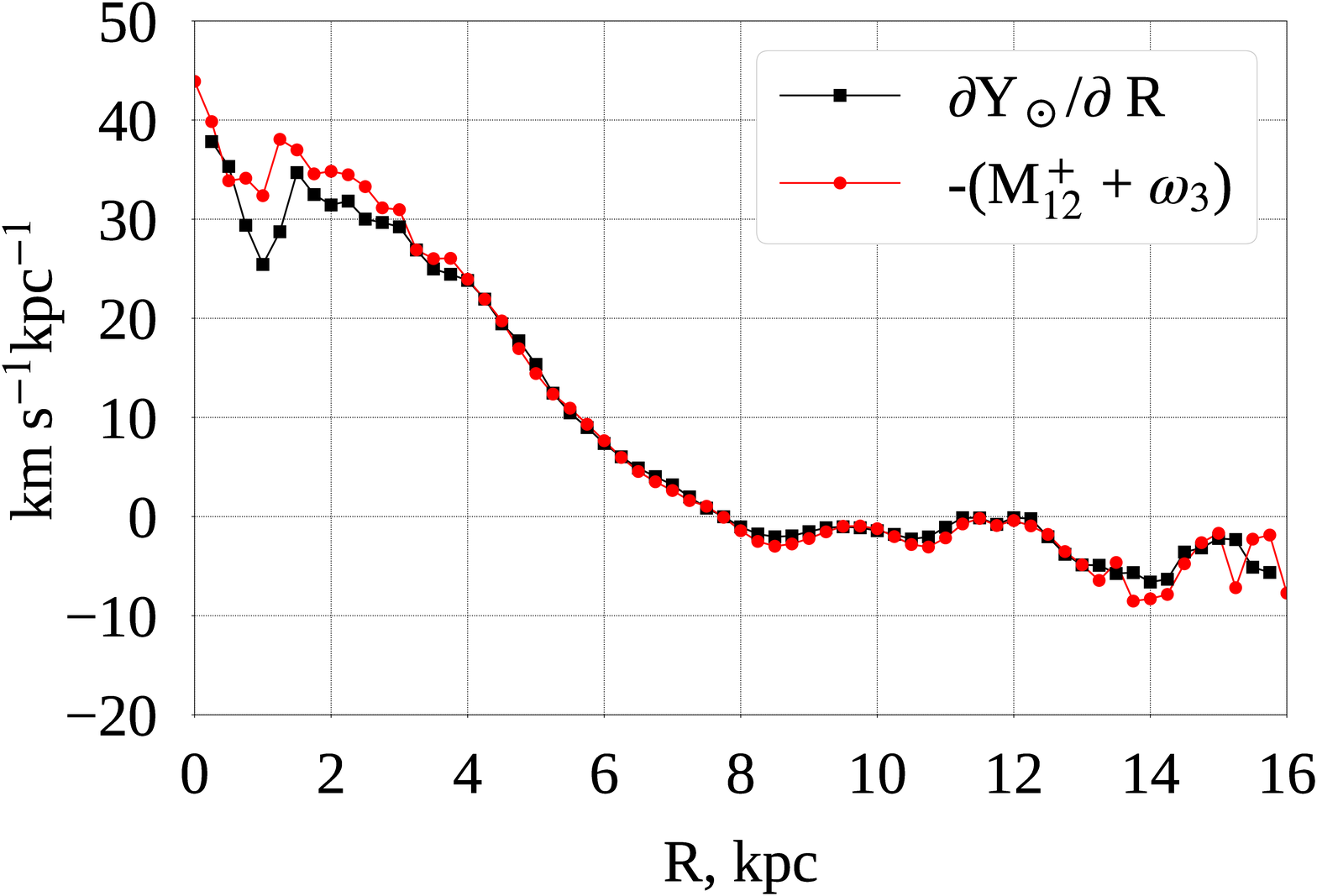}}
  \caption{Comparison of the slope of the Galactic rotational curve $-(M^+_{12}+\omega_3)$ and the numerical derivative $\partial Y_\odot/\partial R$.}
\label{fig:slope-dY}
\end{figure}

At the same time, we can derive the slope of the Galactic rotational curve via differentiation of $Y_\odot$ by $R$.
For comparison, Fig.~\ref{fig:slope-dY} demonstrates the numerical derivative of the function $Y_\odot$ by $R$ which is the slope of the Galactic rotational velocity derived only from the analysis of the peculiar Solar velocity relative to the centroids. Its values are in good agreement with those computed from the relation \ref{eq:Vrot_slope}.

\subsection {Vertical gradient of the Galactic rotational curve}

The vertical gradient of the Galactic rotational velocity $\partial V_{\rm rot}/\partial z$ is defined as follows:
\begin{equation}
\label{eq:gal_grad}
\frac{\partial V_{\rm rot}}{\partial z} = M^+_{23} - \omega_1
\end{equation}
The analysis of the kinematic parameter using vector spherical harmonics was carefully carried out in the papers by \cite{Vityazev2014, Velichko2020}. It was shown that the gradients derived from the $Gaia$~DR2 data in the northern and southern hemispheres have opposite signs and their modules are nearly equal. Therefore, when calculating over the entire sphere, within the distance range 4--12~kpc $\partial V_{\rm rot}/\partial z$ is nearly zero but still non-zero. The fact is given in the left panel of Fig.~\ref{fig:ang_vel} (in the middle) that shows the dependency of $\partial V_{\rm rot}/\partial z$ on $R$.

\subsection {Galactic warp}
The relation
\begin{equation}
\label{eq:gal_warp}
\Sigma = -\frac{1}{R}\frac{\partial V_z}{\partial \theta} = \omega_1 + M^+_{23}
\end{equation}
is usually interpreted as kinematic manifestation of the local Galactic warp. Its values were derived for the Solar vicinity by \cite{Miyamoto1993, Zhu2000, Mignard2000, Vityazev2012}. In the present paper the value is found to be nearly zero within the entire distance range (the right panel of Fig.~\ref{fig:ang_vel}, in the middle).

\subsection {Gradient of the vertical velocity component along the radius-vector}

The gradient of the vertical component of the stellar velocity field along the radius-vector is defined as:
\begin{equation}
\label{eq:dvz_dr}
\frac{\partial V_z}{\partial R} = \omega_2 - M^+_{13}
\end{equation}
It is shown in the left panel of Fig.~\ref{fig:ang_vel} (bottom raw). One can see that its values are nearly zero within the distance range 2--14~kpc.

\subsection {Vertical gradient of the Galactic expansion velocity}

The vertical gradient of the Galactic expansion velocity can be found from the equation:
\begin{equation}
\label{eq:dvr_dz}
\frac{\partial V_R}{\partial z} = -(\omega_2 + M^+_{13})
\end{equation}

Its behaviour is given in the right panel of Fig.~\ref{fig:ang_vel} (bottom raw). One can see that its numerical values are small but statistically significant. Besides, its values grow slowly with distance up to about 11.5~kpc.
It is obvious that more accurate estimations of the given gradients may be established only solving similar tasks for the northern and southern hemispheres separately.

\subsection {Relationships between the diagonal components of the deformation tensor and the stellar velocity field components 
$ V_R$, $ V_\theta$ and $V_z$}

\begin{equation}
\label{eq:gal_11}
\frac{\partial V_R}{\partial R} = M^+_{11} 
\end{equation}

\begin{equation}
\label{eq:gal_22}
\frac{V_R}{R} + \frac{1}{R}\frac{\partial V_{\theta}}{\partial \theta} = M^+_{22} 
\end{equation}

\begin{equation}
\label{eq:gal_33}
\frac{\partial V_z}{\partial z} = M^+_{33} 
\end{equation}

\begin{figure}
   \centering
\resizebox{\hsize}{!}
   {\includegraphics{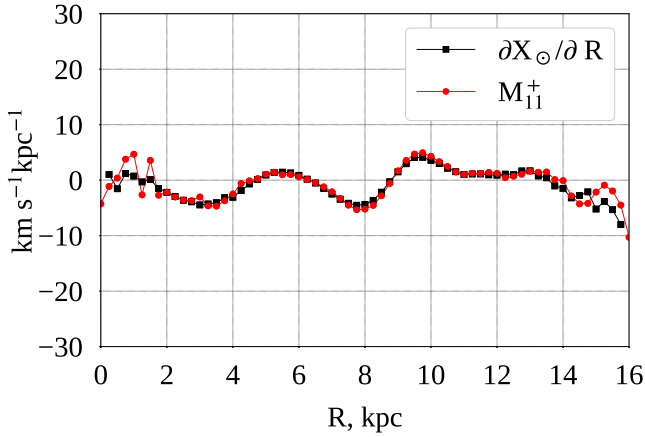}}
  \caption{Comparison of the component of deformation velocity $M^+_{11}$ along axis $x$ and the numerical derivative $\partial X_\odot/\partial R$.}
\label{fig:M11p-dX}%
\end{figure}

As mentioned above, the behaviour of the Solar velocity components $X_\odot, Y_\odot$ and $Z_\odot$ reflects that of centroids' linear velocity components relative the Galactic center. This obviously implies that the centroid's velocity vector has not only the $Y_\odot$ component related to $V_{\theta}$ but $X_\odot$ and $Z_\odot$ components related to $V_{R}$ and $V_{Z}$, respectively. The parameters $X_\odot$ and $M^+_{11}$ shown in the top panel of Fig.~\ref{fig:diagonal_par} have been computed as independent unknowns by solving the O--M model equations. Therefore, the excellent agreement between the $M^+_{11}$ dependency and that derived by numerical differentiation of the function $X_\odot$ by $R$ is an evidence of both actuality and reliability of their estimations (see Fig.~\ref{fig:M11p-dX}).

\section {Summary and conclusions}

The method applied in the present paper for deriving the O--M model kinematic parameters with the use of stellar spatial velocities, allowed both to estimate the parameters in a specific local Galactic region and to compute a number of global characteristics related to much larger Galactic volumes. Besides, this approach have demonstrated appearance of new, alternative possibilities, in particular, to trace the behaviour of the Galactic rotational curve and its slope. The present results demonstrate the ability of the method to check consistency and reliability of the derived parameters. Numerical estimations of the kinematic parameters as well as their errors are given in Tables~\ref{tab:model_part1}, \ref{tab:model_part2} of the Appendix \ref{sec:app1}.
 
Since there are some processes in the Galaxy which have opposite directions in the northern and southern hemispheres, when analysing the stellar velocity field over the entire sphere, they compensate each other and become undetectable. Therefore, the final interpretation of the derived results requires additional investigations of the stellar velocity field in the northern and southern Galactic hemispheres separately. In the near future, we plan to publish the results of further research aimed at eliminating the uncertainties appeared at large and small Galactocentric distances as well as enlarging the volume of the Galactic spatial area under study.

\section{Acknowledgements}
This work has made use of data from the European Space Agency (ESA) mission {\it Gaia} (\url{https://www.cosmos.esa.int/gaia}), processed by the {\it Gaia} Data Processing and Analysis Consortium (DPAC,\url{https://www.cosmos.esa.int/web/gaia/dpac/consortium}). Funding for the DPAC has been provided by national institutions, in particular the institutions participating in the {\it Gaia} Multilateral Agreement.

\section*{Data availability}
\addcontentsline{toc}{section}{Data availability}
The used catalogue data is available in a standardised format for readers via the CDS (https://cds.u-strasbg.fr).
The software code used in this paper can be made available upon request by emailing the corresponding author.

\appendix
\section{Kinematic parameters of the O--M model}
\label{sec:app1}

\begin{table*}
\centering
\caption{The O--M model parameters for stellar systems with $r'$ = 1.0~kpc. $R$ is the Galactocentric distance to centroids, in kpc; N$_{\rm stars}$ is the number of stars containing in each stellar system; the Solar motion components are in \kms; the other O--M model parameters are in \kmskpc.}
\label{tab:model_part1}
\begin{tabular}{llcccccc}
\hline
$R$ &  N$_{\rm stars}$ & $X_\odot$ &  $Y_\odot$  &  $Z_\odot$ &  $\omega_1$ &  $\omega_2$ &  $\omega_3$  \\[0.5mm]
\hline
16.00& 202 & 8.01$\pm$2.02 & 44.98$\pm$2.02 & 5.75$\pm$2.02 & 1.76$\pm$3.24 & -3.47$\pm$3.19 & -2.36$\pm$3.00 \\ 
15.75& 242 & 9.68$\pm$1.78 & 43.85$\pm$1.78 & 7.83$\pm$1.78 & 0.32$\pm$2.79 & -6.32$\pm$2.77 & -6.30$\pm$2.63 \\ 
15.50& 296 & 12.00$\pm$1.57 & 42.16$\pm$1.57 & 6.39$\pm$1.57 & 0.89$\pm$2.50 & -3.02$\pm$2.47 & -5.53$\pm$2.32 \\ 
15.25& 372 & 12.35$\pm$1.33 & 41.30$\pm$1.33 & 4.11$\pm$1.33 & 4.43$\pm$2.09 & 0.10$\pm$2.04 & -3.51$\pm$1.96 \\ 
15.00& 469 & 13.92$\pm$1.30 & 41.00$\pm$1.30 & 4.09$\pm$1.30 & 3.58$\pm$2.01 & 2.59$\pm$2.01 & -7.68$\pm$1.91 \\ 
14.75& 619 & 14.95$\pm$1.18 & 40.19$\pm$1.18 & 4.37$\pm$1.18 & 3.71$\pm$1.78 & 1.37$\pm$1.82 & -7.30$\pm$1.72 \\ 
14.50& 809 & 14.97$\pm$1.04 & 39.41$\pm$1.04 & 3.90$\pm$1.04 & 5.00$\pm$1.58 & -1.09$\pm$1.59 & -4.61$\pm$1.49 \\ 
14.25& 1055 & 16.32$\pm$0.88 & 38.40$\pm$0.88 & 3.82$\pm$0.88 & 4.53$\pm$1.33 & -2.51$\pm$1.34 & -1.81$\pm$1.29 \\ 
14.00& 1383 & 16.56$\pm$0.74 & 36.25$\pm$0.74 & 3.72$\pm$0.74 & 4.38$\pm$1.13 & -1.84$\pm$1.13 & -1.99$\pm$1.07 \\ 
13.75& 1815 & 17.05$\pm$0.69 & 35.10$\pm$0.69 & 3.64$\pm$0.69 & 5.53$\pm$1.04 & -1.37$\pm$1.04 & -1.46$\pm$0.98 \\ 
13.50& 2396 & 17.08$\pm$0.63 & 33.42$\pm$0.63 & 3.47$\pm$0.63 & 6.57$\pm$0.94 & -1.31$\pm$0.94 & -4.09$\pm$0.88 \\ 
13.25& 3160 & 16.84$\pm$0.53 & 32.23$\pm$0.53 & 3.18$\pm$0.53 & 6.99$\pm$0.80 & -0.73$\pm$0.80 & -3.47$\pm$0.75 \\ 
13.00& 4183 & 16.70$\pm$0.45 & 30.97$\pm$0.45 & 2.96$\pm$0.45 & 5.76$\pm$0.69 & -0.71$\pm$0.69 & -4.92$\pm$0.64 \\ 
12.75& 5438 & 16.00$\pm$0.39 & 29.79$\pm$0.39 & 2.88$\pm$0.39 & 5.30$\pm$0.59 & -0.33$\pm$0.59 & -6.11$\pm$0.55 \\ 
12.50& 7209 & 15.87$\pm$0.32 & 29.07$\pm$0.32 & 3.09$\pm$0.32 & 4.26$\pm$0.49 & -0.30$\pm$0.50 & -7.40$\pm$0.45 \\ 
12.25& 9410 & 15.50$\pm$0.28 & 28.77$\pm$0.28 & 3.71$\pm$0.28 & 2.82$\pm$0.43 & -0.21$\pm$0.43 & -7.83$\pm$0.39 \\ 
12.00& 12527 & 15.33$\pm$0.25 & 28.96$\pm$0.25 & 4.41$\pm$0.25 & 2.13$\pm$0.38 & -0.86$\pm$0.38 & -8.42$\pm$0.34 \\ 
11.75& 16842 & 15.06$\pm$0.22 & 28.72$\pm$0.22 & 4.75$\pm$0.22 & 1.39$\pm$0.32 & -1.33$\pm$0.33 & -8.30$\pm$0.29 \\ 
11.50& 22936 & 14.85$\pm$0.19 & 28.57$\pm$0.19 & 5.15$\pm$0.19 & 0.69$\pm$0.28 & -1.88$\pm$0.28 & -9.41$\pm$0.26 \\ 
11.25& 30955 & 14.49$\pm$0.17 & 28.65$\pm$0.17 & 5.48$\pm$0.17 & 0.54$\pm$0.25 & -2.15$\pm$0.25 & -9.73$\pm$0.22 \\ 
11.00& 42149 & 14.29$\pm$0.14 & 28.50$\pm$0.14 & 5.73$\pm$0.14 & 0.51$\pm$0.21 & -2.03$\pm$0.22 & -9.56$\pm$0.19 \\ 
10.75& 56686 & 13.98$\pm$0.13 & 28.11$\pm$0.13 & 5.92$\pm$0.13 & 0.65$\pm$0.19 & -1.96$\pm$0.20 & -9.49$\pm$0.17 \\ 
10.50& 74710 & 13.54$\pm$0.11 & 27.48$\pm$0.11 & 5.99$\pm$0.11 & 0.76$\pm$0.17 & -1.38$\pm$0.17 & -9.85$\pm$0.15 \\ 
10.25& 96055 & 12.90$\pm$0.10 & 26.98$\pm$0.10 & 6.22$\pm$0.10 & 0.79$\pm$0.15 & -0.85$\pm$0.16 & -10.51$\pm$0.13 \\ 
10.00& 119880 & 12.04$\pm$0.08 & 26.58$\pm$0.08 & 6.56$\pm$0.08 & 0.50$\pm$0.14 & -0.58$\pm$0.14 & -11.22$\pm$0.12 \\ 
9.75& 145752 & 11.12$\pm$0.07 & 26.27$\pm$0.07 & 6.83$\pm$0.07 & 0.33$\pm$0.13 & -0.55$\pm$0.13 & -11.49$\pm$0.11 \\ 
9.50& 172728 & 9.98$\pm$0.07 & 26.01$\pm$0.07 & 7.06$\pm$0.07 & 0.18$\pm$0.12 & -0.60$\pm$0.13 & -11.43$\pm$0.10 \\ 
9.25& 199244 & 9.09$\pm$0.06 & 25.76$\pm$0.06 & 7.26$\pm$0.06 & -0.00$\pm$0.12 & -0.76$\pm$0.12 & -11.22$\pm$0.09 \\ 
9.00& 225976 & 8.51$\pm$0.06 & 25.45$\pm$0.06 & 7.43$\pm$0.06 & -0.21$\pm$0.12 & -0.83$\pm$0.12 & -10.94$\pm$0.09 \\ 
8.75& 250562 & 8.34$\pm$0.06 & 25.00$\pm$0.06 & 7.52$\pm$0.06 & -0.24$\pm$0.11 & -0.85$\pm$0.11 & -11.06$\pm$0.08 \\ 
8.50& 273889 & 8.65$\pm$0.06 & 24.47$\pm$0.06 & 7.63$\pm$0.06 & -0.14$\pm$0.11 & -0.90$\pm$0.11 & -11.47$\pm$0.08 \\ 
8.25& 296300 & 9.43$\pm$0.05 & 23.98$\pm$0.05 & 7.67$\pm$0.05 & -0.10$\pm$0.11 & -0.93$\pm$0.11 & -12.39$\pm$0.08 \\ 
8.00& 320733 & 10.47$\pm$0.05 & 23.59$\pm$0.05 & 7.72$\pm$0.05 & -0.04$\pm$0.11 & -0.91$\pm$0.11 & -13.50$\pm$0.08 \\ 
7.75& 347258 & 11.62$\pm$0.05 & 23.46$\pm$0.05 & 7.79$\pm$0.05 & 0.06$\pm$0.11 & -0.72$\pm$0.11 & -14.66$\pm$0.08 \\ 
7.50& 375367 & 12.73$\pm$0.05 & 23.58$\pm$0.05 & 7.88$\pm$0.05 & 0.07$\pm$0.11 & -0.57$\pm$0.11 & -15.56$\pm$0.08 \\ 
7.25& 401070 & 13.70$\pm$0.05 & 23.88$\pm$0.05 & 7.97$\pm$0.05 & 0.13$\pm$0.10 & -0.61$\pm$0.10 & -16.11$\pm$0.07 \\ 
7.00& 417071 & 14.45$\pm$0.05 & 24.57$\pm$0.05 & 8.09$\pm$0.05 & 0.37$\pm$0.10 & -0.28$\pm$0.10 & -16.83$\pm$0.08 \\ 
6.75& 418459 & 14.95$\pm$0.05 & 25.48$\pm$0.05 & 8.20$\pm$0.05 & 0.61$\pm$0.10 & -0.28$\pm$0.10 & -17.46$\pm$0.08 \\ 
6.50& 405412 & 15.20$\pm$0.05 & 26.58$\pm$0.05 & 8.34$\pm$0.05 & 0.72$\pm$0.11 & -0.10$\pm$0.11 & -18.18$\pm$0.08 \\ 
6.25& 376332 & 15.21$\pm$0.06 & 27.93$\pm$0.06 & 8.48$\pm$0.06 & 0.75$\pm$0.11 & 0.02$\pm$0.11 & -18.99$\pm$0.09 \\ 
6.00& 332420 & 15.10$\pm$0.07 & 29.60$\pm$0.07 & 8.68$\pm$0.07 & 0.69$\pm$0.12 & 0.10$\pm$0.12 & -20.20$\pm$0.10 \\ 
5.75& 276790 & 14.77$\pm$0.08 & 31.62$\pm$0.08 & 8.85$\pm$0.08 & 0.88$\pm$0.13 & 0.24$\pm$0.14 & -21.42$\pm$0.11 \\ 
5.50& 218552 & 14.45$\pm$0.10 & 34.09$\pm$0.10 & 8.92$\pm$0.10 & 0.95$\pm$0.15 & 0.25$\pm$0.16 & -22.72$\pm$0.13 \\ 
5.25& 166997 & 14.06$\pm$0.11 & 36.85$\pm$0.11 & 9.04$\pm$0.11 & 0.39$\pm$0.18 & 0.22$\pm$0.18 & -24.05$\pm$0.16 \\ 
5.00& 126335 & 13.75$\pm$0.14 & 40.31$\pm$0.14 & 9.08$\pm$0.14 & 0.29$\pm$0.21 & -0.17$\pm$0.21 & -25.49$\pm$0.19 \\ 
4.75& 94575 & 13.59$\pm$0.17 & 44.52$\pm$0.17 & 9.14$\pm$0.17 & 0.13$\pm$0.25 & 0.36$\pm$0.25 & -27.51$\pm$0.23 \\ 
4.50& 70530 & 13.68$\pm$0.20 & 49.18$\pm$0.20 & 9.37$\pm$0.20 & -0.20$\pm$0.30 & 0.67$\pm$0.30 & -29.60$\pm$0.28 \\ 
4.25& 53084 & 13.93$\pm$0.25 & 54.24$\pm$0.25 & 9.50$\pm$0.25 & 0.05$\pm$0.35 & 0.98$\pm$0.36 & -31.28$\pm$0.35 \\ 
4.00& 40550 & 14.63$\pm$0.29 & 60.15$\pm$0.29 & 9.45$\pm$0.29 & 0.46$\pm$0.42 & 0.70$\pm$0.43 & -33.29$\pm$0.42 \\ 
3.75& 31430 & 15.48$\pm$0.35 & 66.16$\pm$0.35 & 9.42$\pm$0.35 & -0.07$\pm$0.49 & -0.05$\pm$0.51 & -34.84$\pm$0.51 \\ 
3.50& 24730 & 16.20$\pm$0.41 & 72.37$\pm$0.41 & 9.80$\pm$0.41 & -0.91$\pm$0.58 & -0.29$\pm$0.60 & -36.12$\pm$0.61 \\ 
3.25& 19871 & 17.51$\pm$0.48 & 78.65$\pm$0.48 & 9.61$\pm$0.48 & -1.11$\pm$0.68 & -0.96$\pm$0.70 & -37.04$\pm$0.73 \\ 
3.00& 16167 & 18.35$\pm$0.56 & 85.82$\pm$0.56 & 9.16$\pm$0.56 & -1.07$\pm$0.80 & -1.06$\pm$0.81 & -39.66$\pm$0.86 \\ 
2.75& 13375 & 19.74$\pm$0.66 & 93.26$\pm$0.66 & 8.84$\pm$0.66 & -1.17$\pm$0.93 & -1.05$\pm$0.94 & -40.33$\pm$1.02 \\ 
2.50& 11135 & 20.30$\pm$0.76 & 100.65$\pm$0.76 & 9.12$\pm$0.76 & -2.37$\pm$1.07 & 0.10$\pm$1.08 & -41.04$\pm$1.19 \\ 
2.25& 9481 & 21.55$\pm$0.87 & 108.27$\pm$0.87 & 10.47$\pm$0.87 & -2.93$\pm$1.22 & 1.16$\pm$1.23 & -42.33$\pm$1.36 \\ 
2.00& 8118 & 21.78$\pm$0.99 & 116.56$\pm$0.99 & 10.80$\pm$0.99 & -4.87$\pm$1.39 & 0.23$\pm$1.40 & -43.44$\pm$1.55 \\ 
1.75& 7121 & 22.64$\pm$1.11 & 123.98$\pm$1.11 & 9.62$\pm$1.11 & -6.11$\pm$1.54 & -2.77$\pm$1.55 & -44.74$\pm$1.74 \\ 
1.50& 6084 & 22.53$\pm$1.26 & 132.80$\pm$1.26 & 9.33$\pm$1.26 & -6.41$\pm$1.73 & -3.58$\pm$1.75 & -44.18$\pm$1.97 \\ 
1.25& 5172 & 22.59$\pm$1.46 & 141.34$\pm$1.46 & 9.56$\pm$1.46 & -7.76$\pm$1.99 & -3.27$\pm$1.99 & -45.37$\pm$2.26 \\ 
1.00& 4463 & 22.68$\pm$1.66 & 147.17$\pm$1.66 & 7.37$\pm$1.66 & -10.12$\pm$2.24 & -3.08$\pm$2.21 & -44.29$\pm$2.51 \\ 
0.75& 3869 & 22.22$\pm$1.89 & 154.05$\pm$1.89 & 7.85$\pm$1.89 & -10.87$\pm$2.50 & -0.64$\pm$2.46 & -46.63$\pm$2.80 \\ 
0.50& 3399 & 22.08$\pm$2.09 & 161.86$\pm$2.09 & 10.04$\pm$2.09 & -8.25$\pm$2.75 & 5.02$\pm$2.69 & -46.09$\pm$3.07 \\ 
0.25& 2978 & 22.99$\pm$2.31 & 171.72$\pm$2.31 & 13.92$\pm$2.31 & -6.41$\pm$3.02 & 4.24$\pm$3.01 & -50.13$\pm$3.41 \\ 
0.00& 2663 & 21.57$\pm$2.52 & 180.78$\pm$2.52 & 13.87$\pm$2.52 & -7.30$\pm$3.30 & -0.52$\pm$3.25 & -53.29$\pm$3.71 \\ 
\hline
\end{tabular}
\end{table*}

\begin{table*}
\centering
\caption{The O--M model parameters for stellar systems with $r'$ = 1.0~kpc. $R$ is the Galactocentric distance to centroids, in kpc; N$_{\rm stars}$ is the number of stars containing in each stellar system; the O--M model parameters are in \kmskpc.}
\label{tab:model_part2}
\begin{tabular}{llcccccc}
\hline
$R$ &  N$_{\rm stars}$ & $M^+_{23}$ & $M^+_{13}$ & $M^+_{12}$ & $M^+_{11}$ & $M^+_{22}$ & $M^+_{33}$ \\[0.5mm]
\hline
16.00& 202 & -6.92$\pm$3.24 & -5.18$\pm$3.19 & 10.08$\pm$3.00 & -10.29$\pm$4.18 & -3.88$\pm$4.32 & -0.28$\pm$4.82 \\ 
15.75& 242 & -7.93$\pm$2.79 & 1.40$\pm$2.77 & 8.17$\pm$2.63 & -4.49$\pm$3.67 & -3.11$\pm$3.76 & -1.38$\pm$4.13 \\ 
15.50& 296 & -5.45$\pm$2.50 & 2.03$\pm$2.47 & 7.80$\pm$2.32 & -1.95$\pm$3.25 & -4.07$\pm$3.32 & -2.92$\pm$3.74 \\ 
15.25& 372 & -2.99$\pm$2.09 & 1.28$\pm$2.04 & 10.69$\pm$1.96 & -0.90$\pm$2.71 & -2.61$\pm$2.83 & -7.29$\pm$3.06 \\ 
15.00& 469 & -3.37$\pm$2.01 & -0.89$\pm$2.01 & 9.37$\pm$1.91 & -2.16$\pm$2.70 & 1.64$\pm$2.70 & -8.96$\pm$2.97 \\ 
14.75& 619 & -3.54$\pm$1.78 & -0.67$\pm$1.82 & 9.94$\pm$1.72 & -4.18$\pm$2.50 & -0.23$\pm$2.37 & -6.18$\pm$2.64 \\ 
14.50& 809 & -4.93$\pm$1.58 & -0.33$\pm$1.59 & 9.38$\pm$1.49 & -4.28$\pm$2.14 & -1.40$\pm$2.08 & -4.31$\pm$2.37 \\ 
14.25& 1055 & -4.33$\pm$1.33 & 0.02$\pm$1.34 & 9.66$\pm$1.29 & -2.85$\pm$1.84 & 0.17$\pm$1.80 & -2.48$\pm$1.95 \\ 
14.00& 1383 & -3.97$\pm$1.13 & 0.15$\pm$1.13 & 10.28$\pm$1.07 & -0.09$\pm$1.51 & 0.52$\pm$1.53 & -3.71$\pm$1.67 \\ 
13.75& 1815 & -5.96$\pm$1.04 & 0.11$\pm$1.04 & 9.99$\pm$0.98 & 0.10$\pm$1.37 & -0.07$\pm$1.39 & -4.23$\pm$1.56 \\ 
13.50& 2396 & -5.97$\pm$0.94 & -0.68$\pm$0.94 & 8.72$\pm$0.88 & 1.47$\pm$1.25 & -1.60$\pm$1.25 & -3.65$\pm$1.40 \\ 
13.25& 3160 & -5.69$\pm$0.80 & -0.70$\pm$0.80 & 9.92$\pm$0.75 & 1.43$\pm$1.06 & -1.16$\pm$1.06 & -3.70$\pm$1.20 \\ 
13.00& 4183 & -5.07$\pm$0.69 & -1.02$\pm$0.69 & 9.79$\pm$0.64 & 1.63$\pm$0.91 & -0.99$\pm$0.91 & -4.58$\pm$1.03 \\ 
12.75& 5438 & -4.96$\pm$0.59 & -2.15$\pm$0.59 & 9.64$\pm$0.55 & 1.07$\pm$0.78 & -0.01$\pm$0.77 & -3.79$\pm$0.90 \\ 
12.50& 7209 & -4.10$\pm$0.49 & -2.35$\pm$0.50 & 9.20$\pm$0.45 & 0.72$\pm$0.65 & 0.16$\pm$0.64 & -3.89$\pm$0.75 \\ 
12.25& 9410 & -2.94$\pm$0.43 & -2.36$\pm$0.43 & 8.77$\pm$0.39 & 0.44$\pm$0.56 & 0.56$\pm$0.55 & -3.38$\pm$0.65 \\ 
12.00& 12527 & -2.77$\pm$0.38 & -2.76$\pm$0.38 & 8.82$\pm$0.34 & 1.23$\pm$0.49 & 0.87$\pm$0.48 & -2.69$\pm$0.58 \\ 
11.75& 16842 & -1.93$\pm$0.32 & -3.05$\pm$0.33 & 9.20$\pm$0.29 & 1.41$\pm$0.42 & 0.80$\pm$0.41 & -2.03$\pm$0.50 \\ 
11.50& 22936 & -1.63$\pm$0.28 & -2.86$\pm$0.28 & 9.60$\pm$0.26 & 1.19$\pm$0.37 & 0.94$\pm$0.35 & -1.28$\pm$0.43 \\ 
11.25& 30955 & -1.37$\pm$0.25 & -2.63$\pm$0.25 & 10.48$\pm$0.22 & 1.25$\pm$0.33 & 0.85$\pm$0.31 & -0.65$\pm$0.39 \\ 
11.00& 42149 & -1.08$\pm$0.21 & -2.66$\pm$0.22 & 11.71$\pm$0.19 & 1.03$\pm$0.28 & 0.59$\pm$0.26 & -0.14$\pm$0.34 \\ 
10.75& 56686 & -0.86$\pm$0.19 & -2.49$\pm$0.20 & 12.56$\pm$0.17 & 1.48$\pm$0.25 & 0.42$\pm$0.23 & 0.22$\pm$0.30 \\ 
10.50& 74710 & -0.70$\pm$0.17 & -2.36$\pm$0.17 & 12.66$\pm$0.15 & 2.46$\pm$0.22 & 0.45$\pm$0.20 & 0.26$\pm$0.27 \\ 
10.25& 96055 & -0.65$\pm$0.15 & -2.11$\pm$0.16 & 12.53$\pm$0.13 & 3.34$\pm$0.19 & 0.18$\pm$0.18 & -0.02$\pm$0.25 \\ 
10.00& 119880 & -0.40$\pm$0.14 & -2.00$\pm$0.14 & 12.44$\pm$0.12 & 4.30$\pm$0.17 & -0.09$\pm$0.16 & -0.10$\pm$0.23 \\ 
9.75& 145752 & -0.27$\pm$0.13 & -1.60$\pm$0.13 & 12.44$\pm$0.11 & 4.94$\pm$0.16 & -0.60$\pm$0.15 & -0.48$\pm$0.22 \\ 
9.50& 172728 & -0.10$\pm$0.12 & -1.42$\pm$0.13 & 12.41$\pm$0.10 & 4.70$\pm$0.14 & -0.81$\pm$0.14 & -0.59$\pm$0.21 \\ 
9.25& 199244 & 0.08$\pm$0.12 & -1.41$\pm$0.12 & 12.76$\pm$0.09 & 3.54$\pm$0.13 & -1.06$\pm$0.13 & -0.78$\pm$0.20 \\ 
9.00& 225976 & 0.24$\pm$0.12 & -1.23$\pm$0.12 & 13.15$\pm$0.09 & 1.68$\pm$0.13 & -1.26$\pm$0.12 & -0.85$\pm$0.20 \\ 
8.75& 250562 & 0.33$\pm$0.11 & -1.06$\pm$0.11 & 13.82$\pm$0.08 & -0.59$\pm$0.12 & -1.11$\pm$0.12 & -0.84$\pm$0.19 \\ 
8.50& 273889 & 0.25$\pm$0.11 & -1.08$\pm$0.11 & 14.45$\pm$0.08 & -2.81$\pm$0.12 & -0.82$\pm$0.11 & -0.68$\pm$0.19 \\ 
8.25& 296300 & 0.31$\pm$0.11 & -1.07$\pm$0.11 & 14.90$\pm$0.08 & -4.54$\pm$0.11 & -0.37$\pm$0.11 & -0.60$\pm$0.19 \\ 
8.00& 320733 & 0.20$\pm$0.11 & -0.99$\pm$0.11 & 14.92$\pm$0.08 & -5.22$\pm$0.11 & -0.23$\pm$0.11 & -0.38$\pm$0.19 \\ 
7.75& 347258 & -0.04$\pm$0.11 & -0.95$\pm$0.11 & 14.73$\pm$0.08 & -5.33$\pm$0.11 & -0.11$\pm$0.11 & -0.39$\pm$0.19 \\ 
7.50& 375367 & -0.19$\pm$0.11 & -0.88$\pm$0.11 & 14.52$\pm$0.08 & -4.50$\pm$0.11 & -0.12$\pm$0.11 & -0.13$\pm$0.18 \\ 
7.25& 401070 & -0.26$\pm$0.10 & -1.00$\pm$0.10 & 14.50$\pm$0.07 & -3.29$\pm$0.11 & -0.14$\pm$0.11 & 0.21$\pm$0.18 \\ 
7.00& 417071 & -0.47$\pm$0.10 & -0.79$\pm$0.10 & 14.20$\pm$0.08 & -2.11$\pm$0.11 & 0.07$\pm$0.11 & 0.27$\pm$0.18 \\ 
6.75& 418459 & -0.64$\pm$0.10 & -0.84$\pm$0.10 & 13.95$\pm$0.08 & -1.21$\pm$0.11 & 0.36$\pm$0.11 & 0.40$\pm$0.18 \\ 
6.50& 405412 & -0.69$\pm$0.11 & -0.84$\pm$0.11 & 13.64$\pm$0.08 & -0.46$\pm$0.12 & 0.79$\pm$0.11 & 0.37$\pm$0.18 \\ 
6.25& 376332 & -0.67$\pm$0.11 & -0.69$\pm$0.11 & 13.02$\pm$0.09 & 0.12$\pm$0.13 & 1.52$\pm$0.12 & 0.53$\pm$0.19 \\ 
6.00& 332420 & -0.55$\pm$0.12 & -0.71$\pm$0.12 & 12.55$\pm$0.10 & 0.61$\pm$0.14 & 2.35$\pm$0.13 & 0.70$\pm$0.20 \\ 
5.75& 276790 & -0.67$\pm$0.13 & -0.67$\pm$0.14 & 12.11$\pm$0.11 & 1.02$\pm$0.16 & 3.28$\pm$0.15 & 0.98$\pm$0.22 \\ 
5.50& 218552 & -0.81$\pm$0.15 & -0.42$\pm$0.16 & 11.79$\pm$0.13 & 0.98$\pm$0.19 & 3.84$\pm$0.18 & 1.11$\pm$0.25 \\ 
5.25& 166997 & -0.53$\pm$0.18 & -0.22$\pm$0.18 & 11.69$\pm$0.16 & 1.44$\pm$0.22 & 4.19$\pm$0.22 & 1.55$\pm$0.28 \\ 
5.00& 126335 & -0.58$\pm$0.21 & -0.12$\pm$0.21 & 11.06$\pm$0.19 & 1.06$\pm$0.27 & 4.70$\pm$0.27 & 1.55$\pm$0.33 \\ 
4.75& 94575 & -0.80$\pm$0.25 & 0.08$\pm$0.25 & 10.57$\pm$0.23 & 0.27$\pm$0.33 & 4.26$\pm$0.33 & 1.67$\pm$0.38 \\ 
4.50& 70530 & -0.82$\pm$0.30 & 0.46$\pm$0.30 & 9.86$\pm$0.28 & -0.13$\pm$0.40 & 3.30$\pm$0.40 & 1.40$\pm$0.44 \\ 
4.25& 53084 & -0.76$\pm$0.35 & 0.44$\pm$0.36 & 9.34$\pm$0.35 & -0.62$\pm$0.50 & 2.34$\pm$0.49 & 0.65$\pm$0.51 \\ 
4.00& 40550 & -1.07$\pm$0.42 & 0.22$\pm$0.43 & 9.35$\pm$0.42 & -2.50$\pm$0.61 & 0.57$\pm$0.59 & 0.06$\pm$0.60 \\ 
3.75& 31430 & -0.75$\pm$0.49 & 0.26$\pm$0.51 & 8.79$\pm$0.51 & -3.71$\pm$0.74 & -0.50$\pm$0.70 & -1.20$\pm$0.69 \\ 
3.50& 24730 & 0.24$\pm$0.58 & 0.02$\pm$0.60 & 10.10$\pm$0.61 & -4.69$\pm$0.89 & -1.53$\pm$0.85 & -1.67$\pm$0.80 \\ 
3.25& 19871 & 0.32$\pm$0.68 & -0.85$\pm$0.70 & 10.14$\pm$0.73 & -4.62$\pm$1.05 & -1.52$\pm$1.01 & -0.37$\pm$0.91 \\ 
3.00& 16167 & 2.00$\pm$0.80 & -1.52$\pm$0.81 & 8.70$\pm$0.86 & -3.03$\pm$1.24 & -2.84$\pm$1.20 & 0.65$\pm$1.05 \\ 
2.75& 13375 & 3.67$\pm$0.93 & -0.42$\pm$0.94 & 9.19$\pm$1.02 & -3.70$\pm$1.45 & -4.84$\pm$1.42 & 0.50$\pm$1.19 \\ 
2.50& 11135 & 4.57$\pm$1.07 & -0.08$\pm$1.08 & 7.76$\pm$1.19 & -3.57$\pm$1.69 & -5.06$\pm$1.66 & 0.85$\pm$1.35 \\ 
2.25& 9481 & 5.87$\pm$1.22 & -1.78$\pm$1.23 & 7.85$\pm$1.36 & -3.04$\pm$1.93 & -6.05$\pm$1.91 & 1.34$\pm$1.52 \\ 
2.00& 8118 & 6.80$\pm$1.39 & -2.56$\pm$1.40 & 8.61$\pm$1.55 & -2.18$\pm$2.21 & -5.08$\pm$2.18 & 1.82$\pm$1.71 \\ 
1.75& 7121 & 7.40$\pm$1.54 & -0.91$\pm$1.55 & 10.17$\pm$1.74 & -2.72$\pm$2.48 & -7.38$\pm$2.45 & 2.96$\pm$1.87 \\ 
1.50& 6084 & 6.47$\pm$1.73 & -0.13$\pm$1.75 & 7.19$\pm$1.97 & 3.56$\pm$2.80 & -6.93$\pm$2.77 & 1.86$\pm$2.09 \\ 
1.25& 5172 & 7.30$\pm$1.99 & 0.35$\pm$1.99 & 7.31$\pm$2.26 & -2.66$\pm$3.20 & -4.45$\pm$3.20 & 3.91$\pm$2.37 \\ 
1.00& 4463 & 6.80$\pm$2.24 & 2.52$\pm$2.21 & 11.92$\pm$2.51 & 4.67$\pm$3.52 & -9.54$\pm$3.59 & 5.17$\pm$2.67 \\ 
0.75& 3869 & 5.53$\pm$2.50 & -0.96$\pm$2.46 & 12.49$\pm$2.80 & 3.77$\pm$3.92 & -9.24$\pm$4.01 & 8.42$\pm$2.98 \\ 
0.50& 3399 & 5.19$\pm$2.75 & -6.60$\pm$2.69 & 12.22$\pm$3.07 & 0.38$\pm$4.26 & -8.55$\pm$4.43 & 7.43$\pm$3.28 \\ 
0.25& 2978 & 10.07$\pm$3.02 & -6.11$\pm$3.01 & 10.28$\pm$3.41 & -1.12$\pm$4.82 & -2.53$\pm$4.83 & 6.17$\pm$3.62 \\ 
0.00& 2663 & 10.66$\pm$3.30 & -2.32$\pm$3.25 & 9.39$\pm$3.71 & -4.27$\pm$5.18 & -2.13$\pm$5.31 & 5.78$\pm$3.92 \\ 
\hline
\end{tabular}
\end{table*}

\bsp	
\label{lastpage}
\end{document}